\documentclass[sigplan,10pt,screen=false]{acmart}
\usepackage{comment}
\usepackage{endnotes}
\usepackage{graphicx}
\usepackage{tabularx}
\usepackage{arydshln}
\usepackage{multirow, booktabs}
\usepackage{siunitx}
\usepackage{enumerate}
\usepackage{enumitem}
\usepackage{subcaption}
\usepackage{amsmath}
\usepackage[ruled,linesnumbered]{algorithm2e}
\usepackage{listings}
\usepackage[dvipsnames]{xcolor}
\usepackage{pifont}

\definecolor{dkgreen}{rgb}{0,0.6,0}
\definecolor{gray}{rgb}{0.5,0.5,0.5}
\definecolor{mauve}{rgb}{0.58,0,0.82}
\definecolor{codegreen}{rgb}{0,0.6,0}
\definecolor{codegray}{rgb}{0.5,0.5,0.5}
\definecolor{codepurple}{rgb}{0.58,0,0.82}
\definecolor{backcolour}{rgb}{0.95,0.95,0.92}

\usepackage{makecell}

\usepackage{framed}

\usepackage{hyperref}
\usepackage{cleveref}%needs to be loaded *after* hyperref
\crefname{section}{\S}{\SS}
\crefformat{section}{\S#2#1#3} % see manual of cleveref, section 8.2.1
\crefformat{subsection}{\S#2#1#3}
\crefformat{subsubsection}{\S#2#1#3}

\setlist{itemsep=0pt,parsep=0pt,topsep=0pt}% more compact lists

\usepackage[font=small,labelfont=bf,aboveskip=8pt]{caption}
\usepackage[all]{nowidow}

\usepackage{tikz}
\usetikzlibrary{tikzmark, calc, positioning}
\newcommand{\highlight}[2]{\colorbox{#1!17}{$\displaystyle #2$}}
\usepackage[most]{tcolorbox}
\definecolor{light-gray}{gray}{0.95}
\newtcolorbox{observation}{
    center,
    width=\linewidth,
    colframe=light-gray,
    colback=light-gray
}

\usepackage{wrapfig}

\definecolor{americanrose}{rgb}{1.0, 0.01, 0.24}
\definecolor{antiquefuchsia}{rgb}{0.57, 0.36, 0.51}
\definecolor{cadmiumgreen}{rgb}{0.0, 0.42, 0.24}
\definecolor{check}{RGB}{34, 47, 132}
\definecolor{bostonuniversityred}{rgb}{0.8, 0.0, 0.0}
\definecolor{amber(sae/ece)}{rgb}{1.0, 0.49, 0.0}

\newcommand*\circled[1]{\tikz[baseline=(char.base)]{
            \node[shape=circle,fill,inner sep=1pt] (char) {\textcolor{white}{#1}};}}

\newcommand{\system}{\textsf{TensorDex}\xspace}

\newcommand{\tensorid}{\textsf{TensorID}\xspace}
\newcommand{\tensorsketch}{\textsf{TensorSketch}\xspace}
\newcommand{\flexsplit}{\textsf{FlexSplit}\xspace}

\newcommand{\tensorx}{\textsf{TensorX}\xspace}
\newcommand{\tensorpred}{\textsf{TensorPred}\xspace}

\newcommand{\bitx}{\textsf{BitX}\xspace}

\newcommand{\tensordedup}{\textsf{TensorDedup}\xspace} 
 
\newcommand{\salgo}{\textsf{FlexSplit}\xspace}

\newcommand{\fmplus}{\textsf{FM++}\xspace}

\newcommand{\jason}[1]{\textcolor{orange}{Jason: #1}}

\newcommand{\yuec}[1]{{\color{bostonuniversityred}Yue: #1}}
\newcommand{\added}[1]{#1}
\newcommand{\tf}[1]{\noindent\textcolor{purple}{ tf: #1}}

\newcommand{\phead}[1]{\noindent\textbf{#1}} % general paragraph heading 
\newcommand{\fref}[1]{Fig.~\ref{#1}}
\newcommand{\tref}[1]{Table~\ref{#1}}

\newcommand{\sref}[1]{\cref{#1}}
\newcommand{\aref}[1]{Alg.~\ref{#1}}

\newcommand{\clabel}[1]{\phantomsection\label{challenge:#1}\textbf{#1}}
\newcommand{\clink}[1]{\hyperref[challenge:#1]{\textbf{#1}}}

\newcommand{\ImplBox}[1]{
\vspace{-2pt}
\begin{tcolorbox}[breakable,width=0.48\textwidth,title={Implications},boxrule=.3mm,colback=white,coltitle=white,left=.5mm, right=.5mm, top=.5mm, bottom=.5mm]
\textit{#1}
\end{tcolorbox}
}

\newif\ifcrfinal
\crfinaltrue                       % Final version: disable revision colors
\definecolor{revcolor}{rgb}{0.00,0.35,0.72}   % revision blue
\newcommand{\rev}[1]{#1}
\newenvironment{revblock}{\par}{\par}
\newcommand{\refined}[1]{#1}

  {\par}

\definecolor{draftorange}{rgb}{0.85, 0.45, 0.0}
\newcommand{\draft}[1]{#1}

\title{{\system}: A Compact, Tensor-Centric Storage System for Modern AI Models}

\author{Tingfeng Lan}
\affiliation{%
  \institution{University of Virginia}
  \city{Charlottesville}
  \state{VA}
  \country{USA}
}
\email{erc8gx@virginia.edu}

\author{Zirui Wang}
\affiliation{%
  \institution{University of Virginia}
  \city{Charlottesville}
  \state{VA}
  \country{USA}
}
\email{eeb9sd@virginia.edu}

\author{Yunjia Zheng}
\affiliation{%
  \institution{Harvard University}
  \city{Cambridge}
  \state{MA}
  \country{USA}
}
\email{yunjia_zheng@g.harvard.edu}

\author{Zhaoyuan Su}
\affiliation{%
  \institution{University of Virginia}
  \city{Charlottesville}
  \state{VA}
  \country{USA}
}
\email{acf7ea@virginia.edu}

\author{Juncheng Yang}
\affiliation{%
  \institution{Harvard University}
  \city{Cambridge}
  \state{MA}
  \country{USA}
}
\email{juncheng@seas.harvard.edu}

\author{Yue Cheng}
\affiliation{%
  \institution{University of Virginia}
  \city{Charlottesville}
  \state{VA}
  \country{USA}
}
\email{mrz7dp@virginia.edu}

\copyrightyear{2026}
\acmYear{2026}
\setcopyright{cc}
\setcctype{by}
\acmConference[SOSP '26]{ACM SIGOPS 32nd Symposium on Operating Systems Principles}{September 29-October 02, 2026}{Prague, Czech Republic}
\acmBooktitle{ACM SIGOPS 32nd Symposium on Operating Systems Principles (SOSP '26), September 29-October 02, 2026, Prague, Czech Republic}
\acmDOI{10.1145/3830418.3843872}
\acmISBN{979-8-4007-2585-2/2026/09}

\hypersetup{hidelinks}

\begin{document}
\begin{abstract}

\rev{Modern model hubs store hundreds of petabytes of large language models (LLMs), with fine-tuned variants dominating the storage footprint. These variants contain substantial cross-model redundancy that delta compression can exploit by storing only the difference between a target and a reference model. However, compression effectiveness depends critically on choosing a similar reference. At model-hub scale, this is challenging because model lineage metadata is often missing or unreliable, and different tensors within the same model may be most similar to tensors from different models. Consequently, model-level pairing leaves substantial redundancy unexploited.}

\rev{We present {\system}, a lossless, tensor-centric storage system that performs delta compression at tensor granularity. {\system} decomposes models into tensors, uses compact bit-level fingerprints to predict tensor-pair compressibility, and incrementally organizes tensors into multi-center clusters to select effective bases as new models arrive. Evaluated on 2,890 randomly sampled Hugging Face models, {\system} reduces storage footprint by 70.5\%, 37\% lower than the state-of-the-art design. It achieves 22.9~GB/s compression and 28.4~GB/s decompression throughput, 3.86$\times$ and 1.49$\times$ faster than the next-best system, respectively.}

\end{abstract}

 \maketitle

%sections

\section{Introduction}
\label{sec:intro}

% Large language models (LLMs) have become foundational tools in modern artificial intelligence (AI). With the rapid progress in open-source LLM development~\cite{meta2024llama3, meta2024llama3-2, meta-llama-3.1-8b, jiang2023mistral7b, meta2025llama4}, millions of LLMs are now publicly available through model hubs such as Hugging Face~\cite{huggingface} and TensorFlow Hub~\cite{tensorflowhub}. These platforms support uploads, downloads, and sharing of base models and fine-tuned variants, enabling users to adapt models to diverse downstream tasks with minimal effort. 

% Large language models (LLMs) have become foundational infrastructure across modern AI applications~\cite{meta2024llama3, meta2024llama3-2, meta-llama-3.1-8b, jiang2023mistral7b, meta2025llama4}. As open-source development accelerates, model hubs such as Hugging Face~\cite{huggingface} now host millions of LLMs and are growing at an exponential rate~\cite{zipllm_nsdi26}. A defining characteristic of today's repositories is that fine-tuned derivatives~\cite{HFdocumentation,HFmodelGrowth,laufer2025MLecoSystem}---not base models---dominate both the model count and total storage footprint.
% These derivatives proliferate rapidly as users specialize models for diverse downstream tasks, producing highly heterogeneous collections of closely related but structurally divergent variants. As a result, storage growth is driven not by a few large base models, but by vast families of fine-tuned models that differ subtly yet occupy full model size.

\begin{figure}[t]
    \centering

    \includegraphics[width=0.475\textwidth]{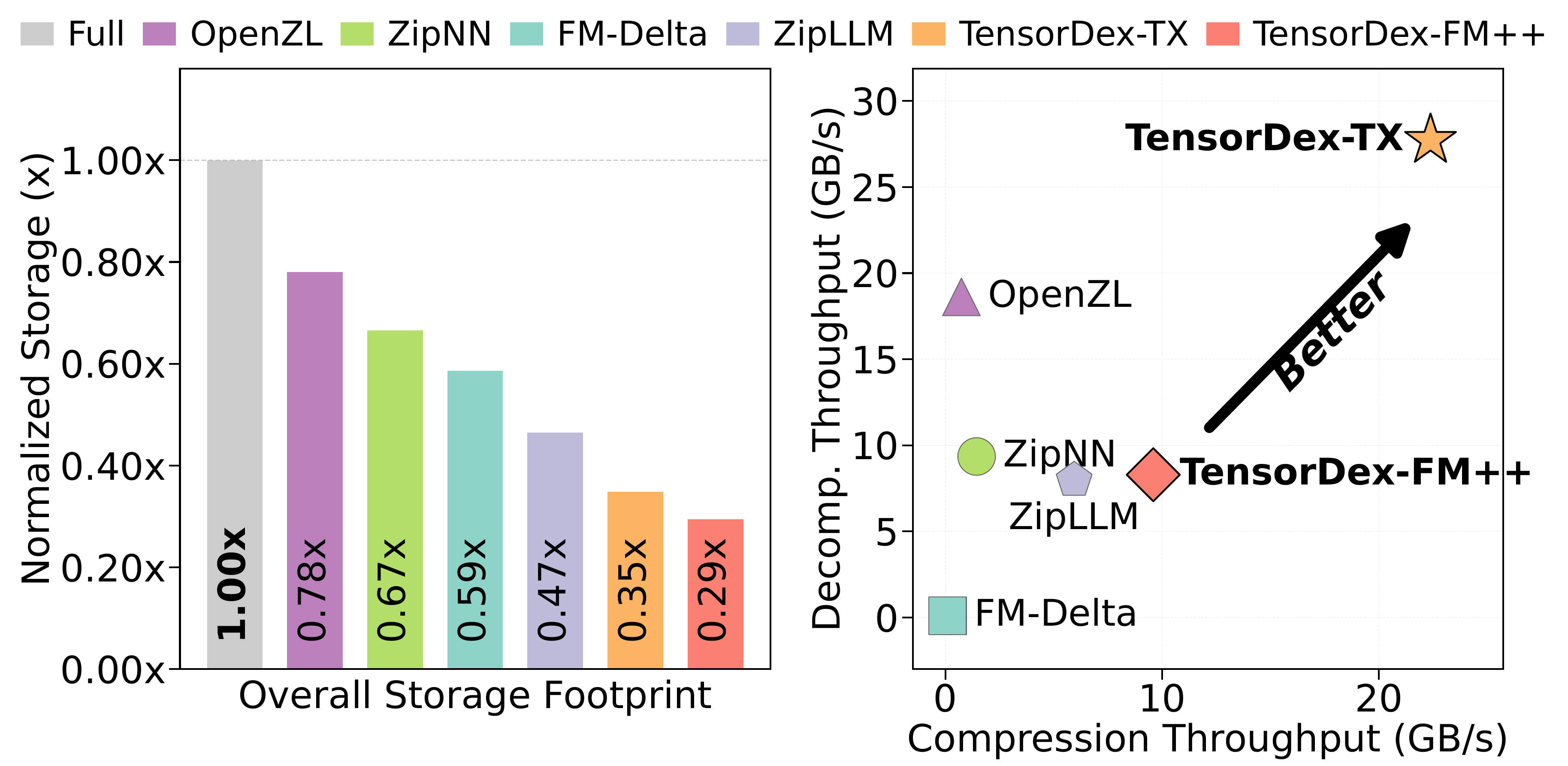}
    \vspace{-15pt}
    \caption{{\bf Left}: Normalized storage relative to uncompressed (Full = 1.00$\times$ 40.11TB of randomly sampled Hugging Face models). {\system} reduces total storage cost by \textbf{3.39$\times$}, achieving substantially lower storage footprint than state-of-the-art baselines. 
    {\bf Right}: {\system} achieves high compression and decompression throughput.} 
    \label{fig:tensordex_key_result}
    %\vspace{-5pt}
\end{figure}

Large language models (LLMs) have become foundational infrastructure across modern AI applications~\cite{meta2024llama3, meta2024llama3-2, meta-llama-3.1-8b, jiang2023mistral7b, meta2025llama4}. As open-source development accelerates, model hubs such as Hugging Face~\cite{huggingface} now host millions of LLMs and are growing at an exponential rate~\cite{zipllm_nsdi26}. 
\rev{The cumulative storage footprint of Hugging Face---models \emph{and} datasets together---has surged from 0.28~PB in 2023 to over 76~PB by the end of 2025~\cite{huggingface_ready_xet_go}. Hugging Face reports that total storage doubled over the following five months and will cross 1~EB by the end of 2026~\cite{hf_storage_exabyte_2026}; within that total, model storage grows faster than dataset storage~\cite{xet_dedup_blog}, so model storage alone is set to reach exabyte scale around 2027.}
%with this growth rate, Hugging Face will surpass the scale of 1~EB model storage by the end of 2027. 
%As shown in \fref{fig:modelhub_growth}, cumulative storage on Hugging Face has surged from 7\,TB in 2021 to over 2,900\,TB by early 2025, with model count rising from 3K to 377K over the same period.\jason{add a projection} \zirui{fig1 and fig2 order?}

\begin{figure}[t]
    \centering

    \includegraphics[width=0.475\textwidth]{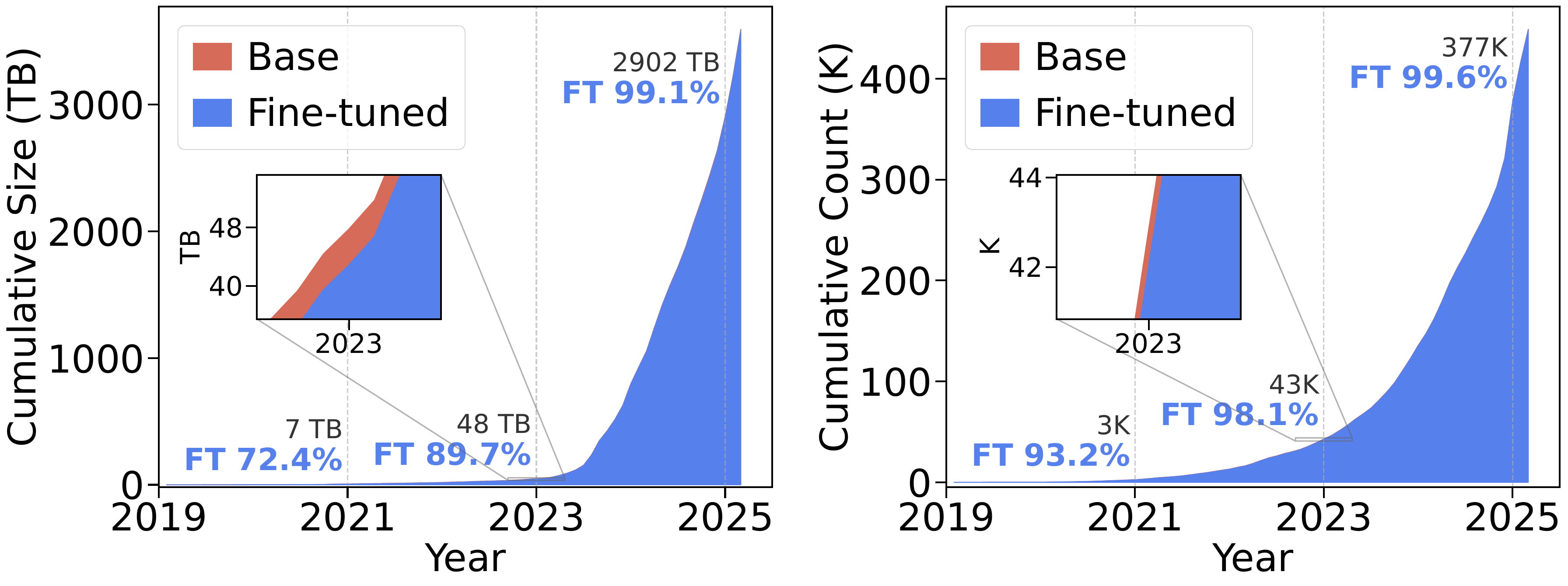}
    \vspace{-15pt} 
    \caption{Cumulative storage size ({\bf left}) and model count ({\bf right}) on Hugging Face from 2019 to 2025. 
    Fine-tuned models (blue) dominate both metrics, accounting for 99.1\% of storage and 99.6\% of model count by 2025, while base models (red) remain a small fraction.}
    \label{fig:modelhub_growth}
    %\vspace{-5pt}
\end{figure}

A defining characteristic of this growth is that fine-tuned derivatives~\cite{HFdocumentation,HFmodelGrowth,laufer2025MLecoSystem}---not base models---dominate both the model count and total storage footprint: by 2025, fine-tuned models account for 99.1\% of total storage and 99.6\% of all models (\fref{fig:modelhub_growth}).
These derivatives proliferate rapidly as users specialize models for diverse downstream tasks, producing highly heterogeneous collections of closely related but structurally divergent variants. As a result, storage growth is driven not by a few large base models, but by vast families of fine-tuned models that differ subtly yet occupy full model size.
This trend is further amplified by model development practices: large-scale training pipelines often retain numerous intermediate checkpoints, each storing full model weights and the optimizer state. For example, recent reports indicate that production training workflows may persist all checkpoints for traceability~\cite{bytecheckpoint_nsdi25}, 
%without aggressive pruning~\cite{bytecheckpoint}, 
significantly compounding the overall storage burden.
%\yuec{HF model storage explosion might not sound too dramatic---people be like, EB-scale, so what. Better add LLM pretraining checkpoint storage accumulation to strengthen the need of model storage reduction.}

Delta compression (i.e., storing only the difference between a fine-tuned model and a reference base) is a natural solution to reduce storage footprint. 
Recent works have demonstrated that it significantly outperform direct compression of model weights~\cite{zipllm_nsdi26, ning2024fm}.
However, deploying delta compression at real-world scale exposes three fundamental limitations of existing approaches:

% old: This emerging landscape fundamentally challenges existing model-aware compression methods such as ZipLLM~\cite{zipllm_nsdi26} \added{and ZipNN~\cite{hershcovitch2024zipnn}}.
% ZipLLM assumes the presence of accurate model-family metadata to pair each fine-tuned model with its base for XOR-style delta compression. However, real-world repositories often lack reliable lineage information: most model repositories at Hugging Face lack model card information; worse, for those with such information, their model cards are user-generated, inconsistently formatted, and frequently incorrect or incomplete. Our analysis on over 500,000 Hugging Face models shows that only 12.4\% contain usable base-model metadata, and even among these, lineage is often wrong. Consequently, metadata-driven compression is infeasible at scale.

\noindent\textbf{Missing Lineage Metadata.}
ZipLLM~\cite{zipllm_nsdi26} relies on Hugging Face model-card metadata to determine model lineage for delta compression. However, model cards are optional. Our analysis of over 2.7M Hugging Face models on March 2026 shows that only 25.8\% contain usable base-model metadata.
When metadata is unavailable, ZipLLM falls back to brute-force pairwise bit distance---computing the exact Hamming distance over model pairs---to infer model similarity. This requires reading every byte of every candidate pair, incurring I/O cost that grows \textit{quadratically} with the size of the model hub, making it impractical for real-world deployment. 
%and is impractical for real-world deployment.

%\noindent\textbf{One Base per Family.} 
% Even if lineage were accurate, the rapid diversification of fine-tuned models renders static, single-center base-delta compression fundamentally inadequate.
\noindent\textbf{Model Pair Sensitivity.}
Even with accurate lineage, assigning all tensors in a fine-tuned model to a single base is fundamentally suboptimal. Fine-tuning induces continuous parameter drift, causing tensors to move progressively farther from any fixed base model.
This drift is amplified by the proliferation of specialized variants: models fine-tuned for reasoning, safety, coding, alignment, or multilingual tasks exhibit significantly different parameter shifts~\cite{gueta2023knowledge,fineTunedModelWeights,weightEnsemble}. 
As a result, compression effectiveness becomes highly sensitive to the choice of base: closely related pairs yield low-entropy deltas and high reduction, while pairing based on a fixed pre-trained base or user-provided metadata often leads to significantly lower compression. 
%As a result, the assumption of a single shared base per model family fails to capture the heterogeneous similarity structure across tensors. 
%, making ``one base per family’’ an unrealistic assumption. 
%\jason{not clear}

%\noindent\textbf{Model-level Granularity.}
% Our deeper analysis reveals an even more structural limitation:
\noindent\textbf{Compression Granularity.}
Model storage redundancy exists at the fine-grained tensor level, not the model level. 
Different tensors within the same model (e.g., embeddings, attention projections, MLP layers) follow distinct training dynamics and thus exhibit different similarity relationships across derivatives. For many tensors, the best bases for delta compression~\cite{zipllm_nsdi26, yao2025deltazip} do not come from the model’s declared base, or even from the same variant; instead, they are scattered across different models.
This mismatch imposes an inherent upper bound on coarse-grained model-level delta compression: forcing all tensors to share a single base leads to systematically suboptimal pairings, limiting achievable reduction ratios even under ideal conditions. 
Model-level compression therefore obscures substantial cross-model redundancy---a coarse, single-base assignment fails to exploit the rich, tensor-level similarity structure present in real-world large-scale model storage. 

% These observations expose a fundamental gap in today’s model storage reduction approaches: 
% \textit{Existing model-aware delta compression methods cannot handle missing metadata (model family) and are not }  
% %\draft{; furthermore, their compression kernels treat delta residuals as generic byte streams, forgoing the substantial additional savings available from exploiting the near-zero, symmetric distribution inherent in LLM weight deltas}. 
% %\draft{As a result, they} fail in the real-world setting where fine-tuned models dominate, diversify, and drift.}
% \added{and therefore fall short in real-world settings where model lineage metadata is often missing or unreliable, and coarse-grained, model-level pairing fails to exploit the rich similarity structure that exists at the tensor level.} 

To this end, we present {\system}, a first-of-its-kind, lossless, \emph{tensor-centric} model storage system that departs from model-level designs and instead \emph{uncovers and exploits hidden cross-tensor relationships across models}, treating tensors as the \emph{first-class citizen} for 
%similarity prediction and 
delta compression. 
%and instead treats tensor as a \emph{first-class citizen} for similarity prediction and delta compression. 

{\system} has four components. 
First, \tensorsketch is a compression-aware fingerprint that captures bit-level structure and enables efficient similarity estimation without accessing full tensor contents. 
Second, a lightweight regressor that predicts the compressibility of tensor pairs using \tensorsketch without reading full tensor data or performing actual compression. 
%materializing actual deltas. 
Third, {\salgo}, a scalable clustering algorithm that identifies the best tensor bases incrementally and adaptively as new models are uploaded. 
%for newly uploaded models. 
Fourth, an optimized codec that performs high-performance compression and decompression. 
Together, these components enable scalable compression planning that adapts to the heterogeneous and evolving structure of real-world model hubs.  

This paper makes the following contributions:

\begin{itemize}[noitemsep,leftmargin=*]
\item This is, to our knowledge, the first work that shows LLM storage redundancy fundamentally emerges at the tensor level, not the model level nor traditional chunk level. 
  
\item We introduce a {\it novel fingerprint metric, \tensorsketch}, to captures both the numerical distribution and coarse structural layout of each tensor, enabling scalable and efficient similarity estimation.

\item We design a regression model to predict model compressibility based on \tensorsketch without reading the full tensor or performing compression. 

\item We design a scalable, incremental clustering algorithm that enable continuous model ingestion by discovering multi-center tensor clusters and selecting high-quality base–delta pairs. % FlexSplit approximates the facility-location objective in fingerprint space, automatically discovering multi-center tensor clusters and selecting high-quality base–delta pairings without relying on metadata. Its incremental nature allows it to naturally adapt to model diversification and drift during continuous ingestion.

\item We implement and optimize two compression codecs that enable high-performance and compression ratio. Together, these components form {\system}. 

\item Evaluated on 2,890 randomly sampled models from Hugging Face, \system reduces storage footprint by 70.5\%, 37\% lower than state-of-the-art design (ZipLLM (\fref{fig:tensordex_key_result})). 
Meanwhile, it achieves 22.9 GB/s compression and 28.4 GB/s decompression throughput, 3.86$\times$ and 1.49$\times$ faster than the next best system, respectively. 

\end{itemize}

\phead{Open-source release.}
\rev{{\system}'s source code is publicly available at 
\url{https://github.com/ds2-lab/TensorDex}.}

\section{Background and Related Work}
\label{sec:background}

\begin{figure}[t]
\centering
\begin{subfigure}{1\linewidth}
  \centering
  \includegraphics[width=\textwidth]{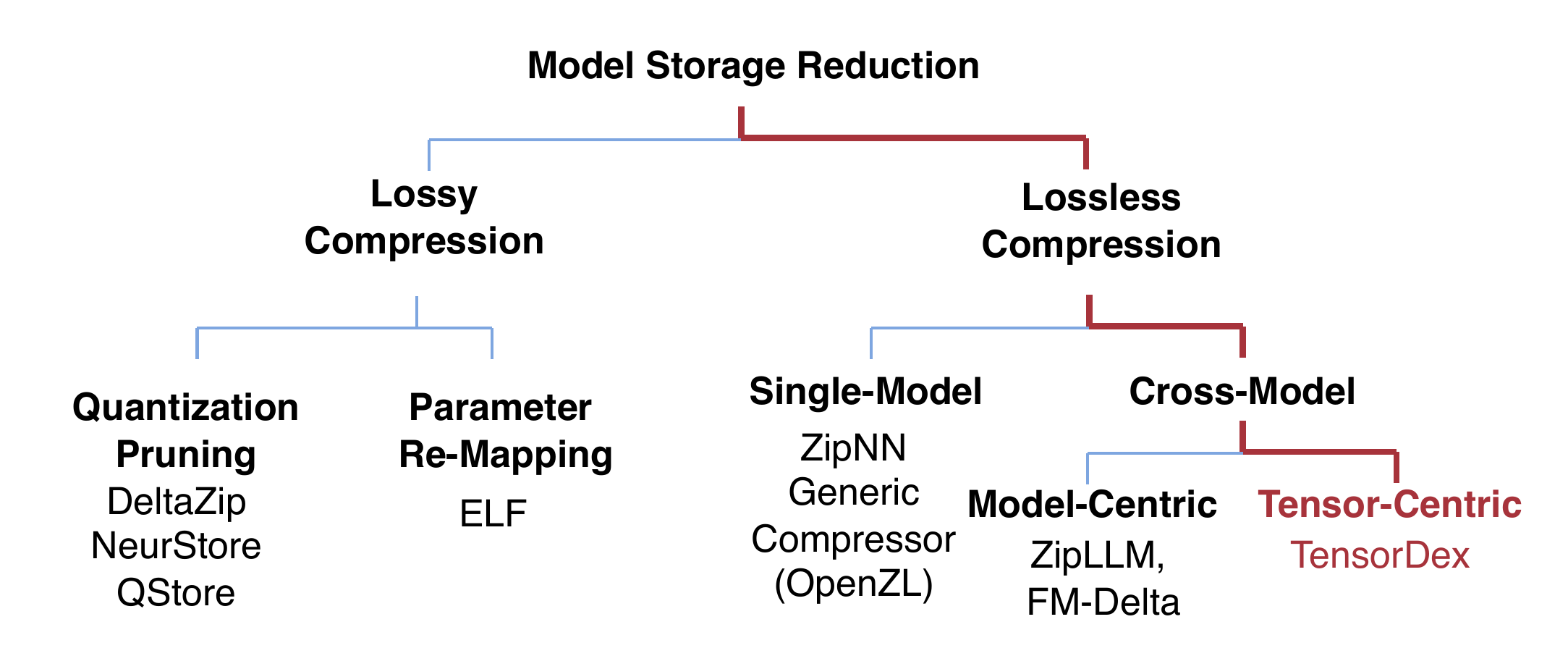}
\end{subfigure}
  \caption{The design scope of \system.} 
\label{fig:related_work}
%\vspace{-5pt}
\end{figure}

This section reviews existing storage reduction techniques.  
%for reducing data storage. 
%including both general-purpose and model-aware approaches. 
A high-level comparison of these
%model storage compression and deduplication 
methods is summarized in ~\fref{fig:related_work}.

\subsection{Traditional Storage Reduction}

\noindent\textbf{General-purpose Compression.}
General-purpose compressors such as Zstandard~\cite{zstd} and Brotli~\cite{Brotli} exploit local byte-level redundancy via dictionary-based~\cite{lz77,lzw} and entropy coding~\cite{huffman,huffman1952method,deutsch1996deflate,duda2013asymmetric,marpe2003context,q_coder_adaptive}, with further gains when the data type is known~\cite{RLH,xdelta,macdonald2000file,kuschewski2023btrblocks,vohra2016practical,liakos2022chimp,blalock2018sprintz,liu2021decomposed,burtscher2007high}; none of them exploits redundancy \emph{across} models. Lossy scientific codecs~\cite{diffenderfer2019error,lindstrom2014fixed,liang2018efficient,di2016fast,tao2017significantly} and inference-time quantization~\cite{lin2024awq,frantar2022gptq,xiao2023smoothquant,zafrir2019q8bert,yao2022zeroquant,dettmers2023spqr} cannot guarantee the exact recovery that model hubs require.

\noindent\textbf{Data Deduplication.} 
Deduplication retains only unique blocks and replaces duplicates with pointers. File-level hashing, as in Git LFS~\cite{git-lfs}, treats a file as atomic, so one changed byte hides all content it still shares with earlier versions. Content-Defined Chunking (CDC)~\cite{LBFS} recovers that redundancy below file granularity using rolling hashes, and is adopted in NetApp ONTAP~\cite{netapp-ontap,netapp-tr3966}, Dell EMC Data Domain~\cite{dell-datadomain}, FastCDC~\cite{fastcdc}, and TiDedup~\cite{TiDedup}. ZipLLM~\cite{zipllm_nsdi26} instead content-hashes each \emph{tensor} and stores it once, capturing exact duplicates across fine-tuned variants that share identical layers, without the metadata overhead of variable-sized chunking.

% \noindent\textbf{General-purpose Deduplication.}
% Deduplication is a widely used technique for reducing storage footprint by identifying and storing only unique data blocks, replacing duplicates with references. File-level deduplication, used in systems like Git LFS~\cite{git-lfs}, eliminates exact file copies with minimal metadata overhead but cannot detect partial redundancy. Chunk-level deduplication, especially content-defined chunking (CDC)~\cite{LBFS}, addresses this by splitting files into variable-sized chunks based on content, enabling more effective duplicate detection despite insertions or shifts. 
% CDC has been adopted in production and research systems such as NetApp ONTAP~\cite{netapp-ontap,netapp-tr3966}, Dell EMC Data Domain~\cite{dell-website,dell-datadomain}, FastCDC~\cite{fastcdc}, and TiDedup~\cite{TiDedup}. 
% However, CDC is ill-suited for model storage due to high metadata overhead from massive variable-sized chunks and limited throughput (50-100~MB/s), making it impractical for processing large-scale model storage efficiently.

%\subsection{ModelHub Evolution}
\subsection{Model-aware Storage Reduction}
\label{sec:bg_model_deduplication}

\phead{Single-model Compression.}
Intra-model compression reduces the storage footprint of individual models by exploiting redundancy in their floating-point parameters.
ELF~\cite{su2024everything} eliminates exponent bits by mapping weights into a normalized range, but it is inherently lossy and unsuitable for model hubs that require exact recovery. ZipNN~\cite{hershcovitch2024zipnn} improves compressibility by reordering float bytes to isolate compressible fields like sign and exponent bits, enabling more effective entropy coding. However, these methods only capture local, single-tensor redundancy and cannot exploit shared patterns across models.

\phead{Cross-model Delta Compression.} 
Delta compression reduces storage by encoding a model’s weights relative to a chosen \emph{base} model. The element-wise difference---by subtraction for floating-point data or XOR for byte-level representations---yields a \emph{delta} of much lower entropy, which a general-purpose compressor then stores. 
FM-Delta~\cite{ning2024fm} compresses arithmetic residuals between base models and their variants, but offers limited throughput and no pairing guidance, making it hard to scale to hubs with millions of models. 
ZipLLM~\cite{zipllm_nsdi26} applies bitwise XOR delta compression to scale storage optimization across large model hubs, \rev{but depends on \texttt{base\_model} metadata and pairs at model granularity (\sref{sec:intro})}.  
DeltaZip~\cite{yao2025deltazip} further prunes the delta matrix with quantization. 
%but its lossy nature makes it unsuitable for model storage that requires exact recovery. 
NeurStore~\cite{neurstore_sigmod25} and QStore~\cite{qstore_vldb26} explore cross-model redundancy via lossy quantization and approximate sharing. In contrast, {\system} preserves exact recovery and targets fine-grained tensor-level redundancy through lossless compression.

\if 0
model-family information on the hub cannot be treated as a dependable signal for large-scale storage planning. Without reliable family information, ZipLLM~\cite{zipllm_nsdi26} defaults back to intra-tensor optimizations (e.g., ZipNN~\cite{hershcovitch2024zipnn}), reducing its overall effectiveness from roughly 50\% storage reduction down to about 30\%.
\fi

\subsection{Vector Similarity Search}
\label{sec:bg_vector_search}
A key in delta compression is to find the most similar pair to compress, which shares similarity with vector search. 
High-dimensional vector search relies on Approximate Nearest Neighbor (ANN) algorithms to address the scaling limits of exact retrieval, including hashing-based~\cite{LSH2004,NNSvision2006,DataDependenthHash2015,LSHcollison2012}, graph-based~\cite{NavigateGraph2019,HVS2021,QueryDrivenGraph2012,smallWorldGraph}, tree-based~\cite{KDtree2008,Trees4NN2006}, and quantization-based~\cite{PQFastScan2015,PQforNNS2011} methods. 
% Early approaches like Locality-Sensitive Hashing (E2LSH)~\cite{LSH2004,NNSvision2006} utilize randomized projections to bucket similar items. 

% They are often paired with Product Quantization (PQ), which compresses vectors into compact subspaces and improves storage efficiency while reducing search latency. 
Graph-based indices currently offer the best balance between speed and accuracy: Hierarchical Navigable Small World (HNSW)~\cite{HNSW2018} routes greedily through a hierarchy of proximity graphs, achieving logarithmic search complexity.
Vector search is often designed for small vectors, e.g., up to 3,072 dimensions; however, tensors can have millions of elements (dimensions). Directly using existing vector search algorithms will be computationally prohibitive.  

\rev{A separate line of work keeps such indexes serviceable under continuous updates rather than rebuilding them: SPFresh~\cite{spfresh_sosp23} splits partitions in place, triggered by partition-size balance, and Quake~\cite{quake_osdi25} re-partitions under a query-latency cost model over partition sizes and access frequencies.}

% \begin{refine}
\section{Problem Formulation and Challenges}
\label{sec:problem_definition}
\subsection{Empirical Insights into LLM Delta Compression}
% \subsection{\added{Empirical Insights into Delta Compression of LLM Weights}}
\label{subsec:insights_challenges}

While prior works~\cite{zipllm_nsdi26, ning2024fm} demonstrate that delta compression can be highly effective in reducing storage footprint, real-world model hubs exhibit more complex and heterogeneous structures than those in controlled experiments.

\begin{figure}[t]
    \centering
    %\vspace{-10pt}
    \includegraphics[width=0.48\textwidth]{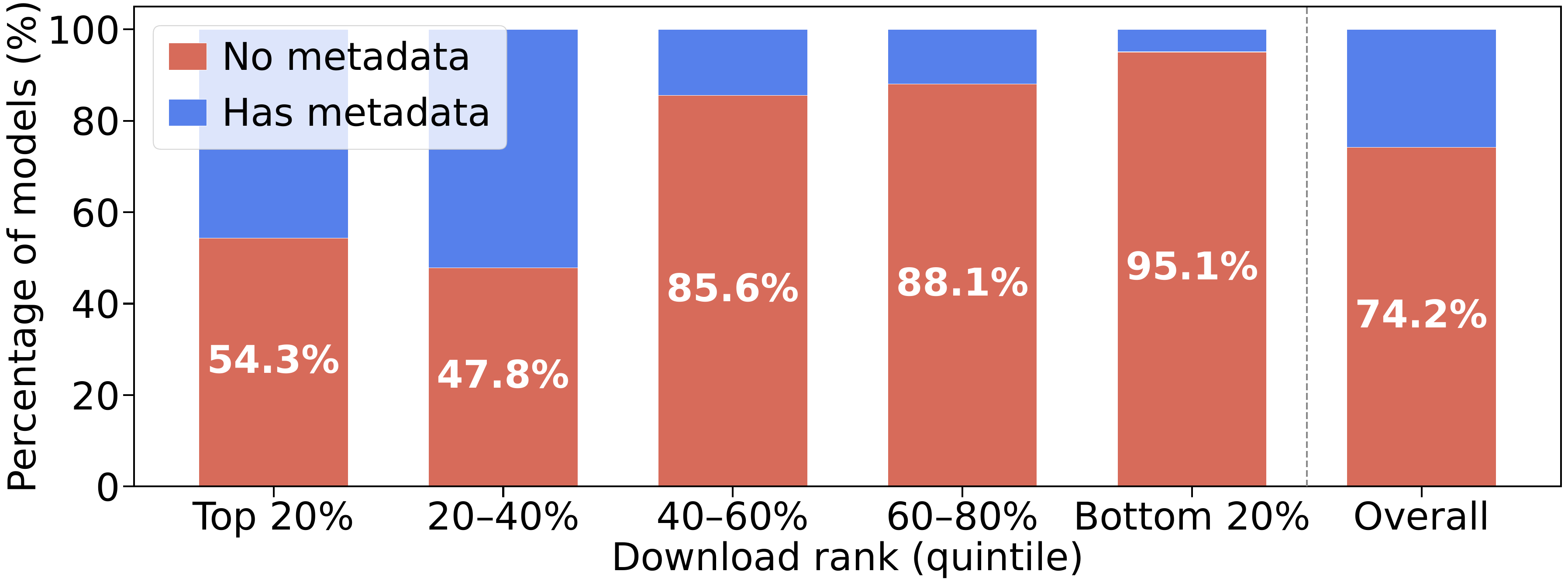}
    \vspace{-15pt}
    \caption{Distribution of Hugging Face model lineage metadata by download rank. Most models (74.2\% overall) lack this information.} 
    %\vspace{-1pt}
    \label{fig:modelhub_metadata}
\end{figure}

% \phead{Missing model lineage information.} 
\phead{Observation \#1 (missing lineage): Model lineage information is often missing or unreliable in real-world model repositories.} 
Existing systems assume model family information is available (e.g., via model cards~\cite{hf-model-cards}); however, such information is often missing or unreliable on the Hugging Face model hub. 
% We find that the majority of the model repositories are inconsistently tagged or entirely unannotated. 
As shown in \fref{fig:modelhub_metadata}, we analyzed all 2,751,212 Hugging Face model repositories as of March 2026 and found that only 25.8\%
%25.83\% 
are tagged with model-family information. 
Even the top 20\% hot models by download count already have a 54.3\% missing rate for lineage metadata. For cold models below the $40^{th}$ percentile in downloads, which make up the majority of the hub, the missing rate consistently exceeds 
88\%. Among those, the metadata is often incomplete or unreliable, as users frequently copy metadata from other repositories or manually modify fields in ways that break the true lineage~\cite{suryani2025model, liang2024s,peft_issue_938}. 

\ImplBox{It is nearly impossible to rely on model-family lineage as a reliable signal for delta compression in real-world model hubs due to missing and unreliable metadata.} 

% Models stored in real-world large-scale model hubs evolve along diverse fine-tuning paths, exhibit uneven similarity across variants, and include many user-generated variants with minimal or inconsistent weight changes, challenging simple-base-delta assumptions. 
% To understand compression behavior in this setting, we perform a systematic study of pairwise delta behavior---using XOR-based delta, as employed by prior work~\cite{zipllm_nsdi26, fmdelta_nips24}---across models and tensors. 
% This study uncovers three key observations about how models evolve on model hubs, capturing complementary aspects of redundancy in model hubs: \emph{pair selection sensitivity} (\#1), 
% %\emph{training-induced drift} (\#2), 
% \emph{compression granularity} (\#2). Together, these observations guide our system design (\cref{sec:design}).

\begin{figure}[ht]
    \centering
    \includegraphics[width=0.48\textwidth]{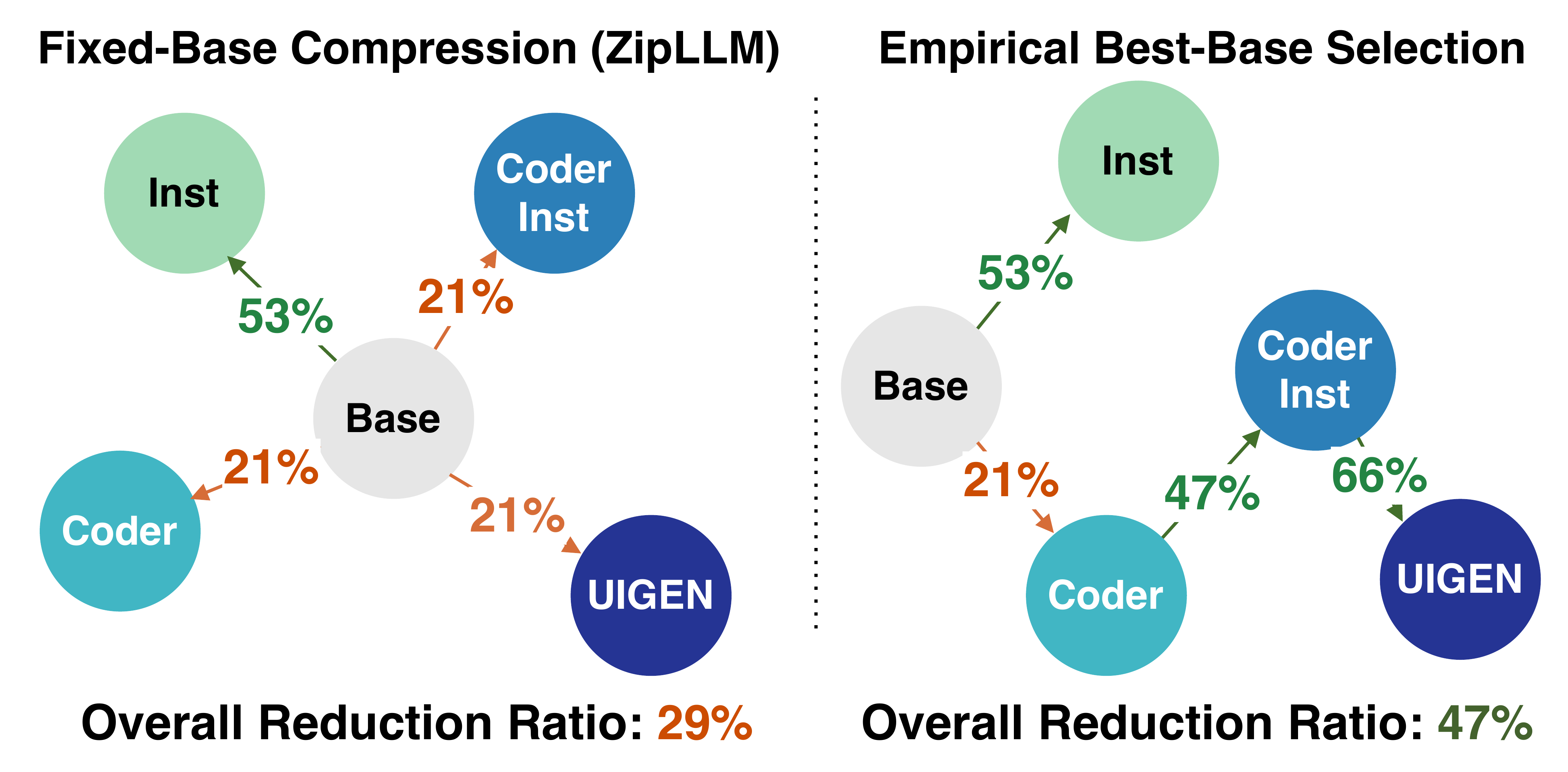}
    \vspace{-15pt}
    \caption{Choosing the delta-compression base is crucial. 
    Both panels show five Qwen2.5-3B variants. Edge labels denote the data-reduction ratio from delta-compressing each variant to its assigned base. 
    \textbf{Left:} Fixed-base compression pairs each variant with the pre-trained \textit{base}, achieving only 29\% overall reduction. \textbf{Right:} Selecting the empirical best base per model raises overall reduction to 47\%.    
    % Both panels show five Qwen2.5-3B variants: the pre-trained \textit{Base}, its \textit{Instruct} and \textit{Coder} fine-tunes, the \textit{Coder-Instruct} variant, and \textit{UIGEN} (for UI generation). Edge labels denote the data reduction ratio achieved by delta-compressing each variant against its assigned base (higher is better). \textbf{Left:} Fixed-base compression (ZipLLM) pairs every variant with the pre-trained \textit{Base}, achieving only 29\% overall reduction—\textit{Coder}, \textit{Coder-Instruct}, and \textit{UIGEN} all compress poorly against the base. \textbf{Right:} Selecting the empirical best base per model raises overall reduction to 47\%, because closely related variants serve as better references for each other (e.g., \textit{Coder-Instruct} → \textit{UIGEN} achieves 64\% vs. only 21\% when paired with the \textit{Base}).
    } 
    \label{fig:optimal_pair}
\end{figure}

%\phead{Observation \#1 (Static view): Delta compression is highly sensitive to pair selection.} 
\phead{Observation \#2 (pair sensitivity): Delta compression is highly sensitive to pair selection.} 
Across LLM repositories, we find that compression effectiveness is governed by the similarity between the target and its base: closely related models yield low-entropy deltas and high reduction, while pairing based on a fixed pre-trained base or user-provided metadata often leads to substantially lower reduction 
%while mismatched pairs produce weak compression 
(\fref{fig:optimal_pair}). 
% Moreover, this sensitivity is amplified by fine-tuning: even within the first 500 training steps, the data reduction ratio (DRR) against the original base drops sharply (e.g., from near 100\% to $\sim$65-70\%), indicating that small weight updates can rapidly degrade compressibility. 
Consequently, pair selection becomes a first-order factor in determining storage efficiency, and naive pairing strategies can leave significant compression potential unrealized.
\ImplBox{Using pre-trained model as the base is insufficient; effective delta compression requires data-driven pairing that reflects true model weight similarity.}

\begin{figure}[ht]
    \centering
 
    \includegraphics[width=0.48\textwidth]{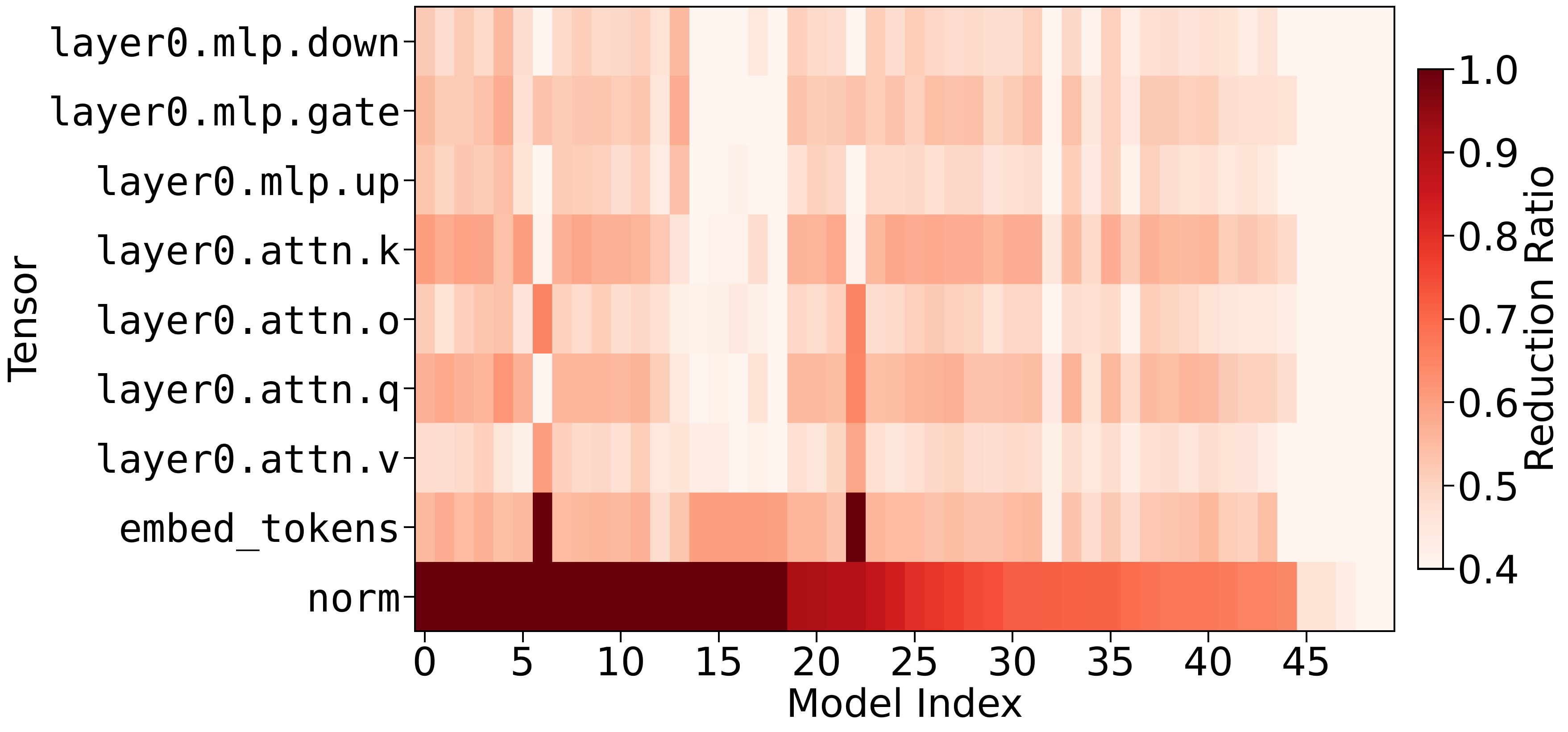}
    \caption{Different tensors within a single model exhibit heterogeneous storage-reduction ratios and the best bases vary.
    Each cell shows the reduction ratio when compressing a tensor (row) against the corresponding tensor in a candidate model (column).}
    \label{fig:compression_matrix_heatmap}
\end{figure}

\if 0
\rev{The uniformly dark \texttt{norm} row is an exception: these small one-dimensional weights compress well against any candidate and contribute negligibly to the storage total.}
\fi

\begin{figure}[ht]
    \centering
    \includegraphics[width=0.48\textwidth]{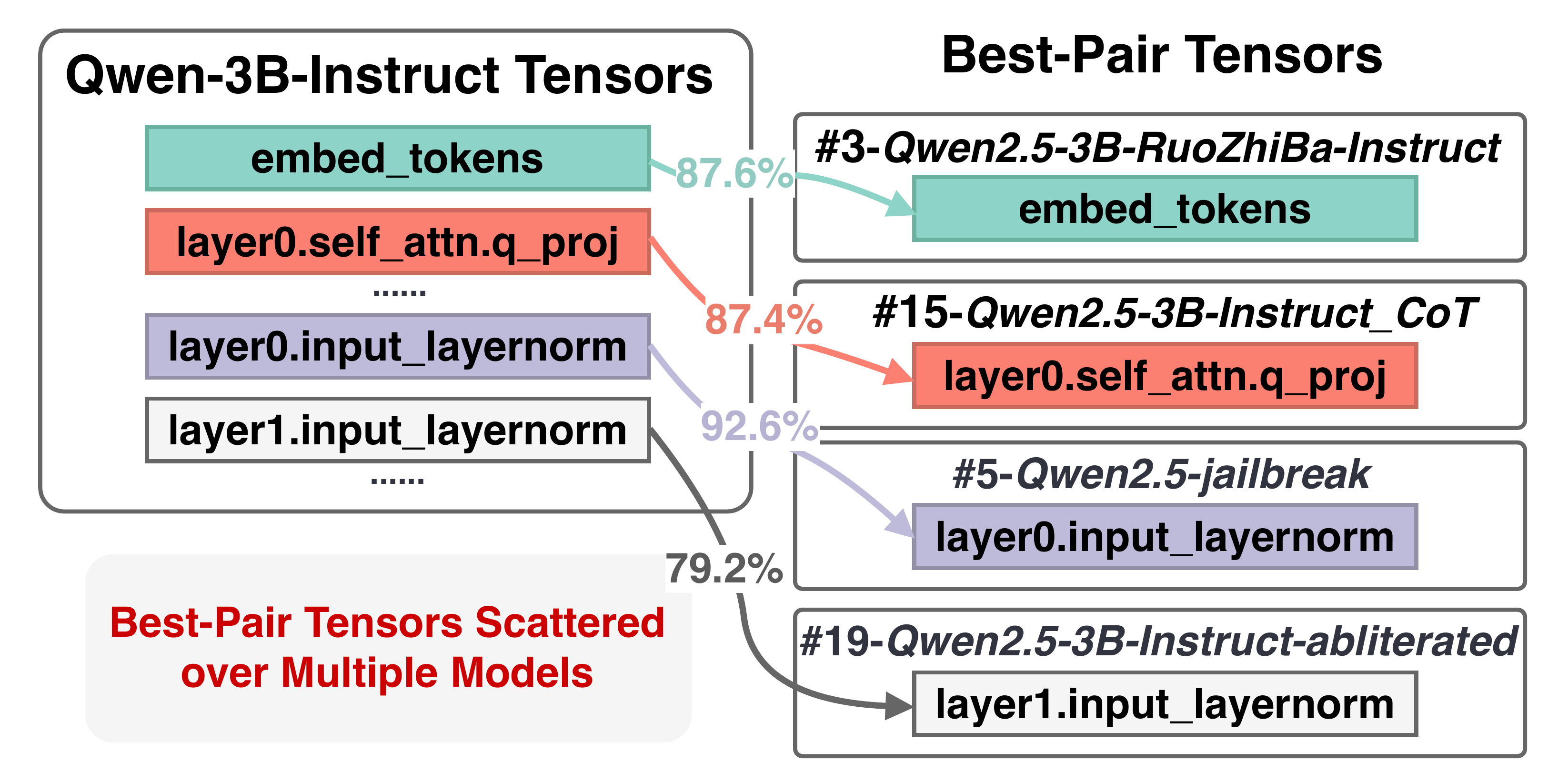}
    \caption{Optimal compression pairing emerges at the tensor level. The best bases for different tensors in the same model often reside in a \emph{different} model, rather than in a single shared base. 
    % For example, the best base tensors for \texttt{embed\_tokens} and \texttt{layer0.self\_attn.q\_proj} come from Model~\#3 (87.6\%) and Model~\#15 (87.4\%), respectively.
    % \textbf{Bottom}: Quantitative evidence for the two highlighted tensors: we plot the reduction ratio against 300 candidate models. The best base tensors for \texttt{embed\_tokens} and \texttt{layer0.self\_attn.q\_proj} come from Model~\#3 (87.6\%) and Model~\#149 (87.4\%), respectively.
    %, while reduction ratios vary widely across candidates confirming the tensor-specific nature of optimal pairing.
    } 
    \label{fig:inter_model}
    %\vspace{-5pt}
\end{figure}

% \begin{figure}[t]
%     \centering
%     \begin{subfigure}[t]{0.235\textwidth}
%         \centering
%         \includegraphics[width=\linewidth]{figures/background/embed_tokens.pdf}
%         \caption{\texttt{embed\_tokens}.}
%         \label{fig:embed_tokens_match}
%     \end{subfigure}
%     \hfill
%     \begin{subfigure}[t]{0.235\textwidth}
%         \centering
%         \includegraphics[width=\linewidth]{figures/background/q_proj_1.pdf}
%         \caption{\texttt{self\_attn.q\_proj}.}
%         \label{fig:q_proj_match}
%     \end{subfigure}
%     \vspace{-6pt}
%     \caption{Compression matches for two representative Qwen-3B-Instruct tensors. For each tensor, we plot its compression ratio against all candidate models. The ``best match model ID’’ denotes the model whose corresponding tensor yields the lowest compression ratio---that is, the most compressible delta. For \texttt{embed\_tokens} (a) and \texttt{self\_attn.q\_proj} (b), their best base tensors come from Model \#5 and Model \#180, respectively.}
%     \label{fig:tensor_pair_compression_ratio}
% \end{figure}

\phead{Observation \#3 (compression granularity): 
The best compression unit is fine-grained---at the tensor level, not the model level.}
%checked
While prior systems typically apply delta encoding at the whole-model granularity~\cite{ning2024fm,zipllm_nsdi26}, our analysis shows that the most effective compression pairing often emerges at the level of individual tensors. Within a single model, different tensors exhibit distinct training dynamics: attention layers, MLP blocks, and embedding matrices drift in different ways as they adapt to data and task objectives. 
\fref{fig:compression_matrix_heatmap} shows that the tensors from a single model may exhibit different data reduction ratios when compressed against a chosen base model (each column). Meanwhile, the best model for each tensor is different (darkest in each row).
Therefore, no single base model consistently serves as the best base for all tensors of a target model. Instead, the optimal pairing varies tensor by tensor, revealing that fine-grained tensor-level decisions are essential for maximizing delta compressibility. 
\fref{fig:inter_model} shows that the best bases of four random tensors from Qwen-3B-Instruct come from four entirely different models. 
% and the reduction ratio varies widely across candidates---illustrating the high variability and tensor-specific nature of optimal pairing.   

%\phead{Observation \#3: The best compression pair is often at the tensor level, not the model level.} While existing systems often perform delta encoding at the whole-model granularity~\cite{ning2024fm,zipllm_nsdi26}, our analysis uncovers that the optimal compression pairing varies across individual tensors, even within the same model. In other words, no single model consistently provides the best XOR partner for all tensors of a target model. This is because different layers or tensor types (e.g., attention weights vs. MLP weights) evolve differently during training, being sensitive to specific data characteristics and task objectives.

\ImplBox{Model storage systems should apply delta compression at the tensor level to better capture cross-model redundancy. Choosing pairs per tensor, rather than per model, improves the data reduction ratio.
%The model storage system should operate at \textbf{tensor granularity} to fully exploit cross-model redundancy. Because the best XOR partner varies from tensor to tensor---and often comes from different models—only per-tensor pairing can reliably capture the most compressible deltas and preserve structural alignment, leading to significantly higher compression efficiency.
}

%Together, the four observations explain why existing systems fail to fully exploit cross-model redundancy:
%compression effectiveness is highly pair-dependent, degrades with training drift, and varies at the tensor rather than model granularity.
%They also make clear that maximizing compression requires deciding \emph{which} tensors should serve as bases and \emph{which} should be stored as compressed deltas. A practical system must therefore jointly address three axes: (i)~\emph{estimating} pairwise tensor similarity efficiently without full data access, (ii)~\emph{predicting} actual compressibility from similarity estimates, and (iii)~\emph{selecting} bases at scale in a dynamic, incremental fashion. This motivates our formal problem definition and the key challenges a practical solution must address.
%and make clear that maximizing compression requires choosing \emph{which} tensors should act as bases and \emph{which} tensors should be compressed and stored as deltas. This motivates our formal problem definition and the key challenges that any practical solution must overcome.

\subsection{Problem Formulation}
\label{subsec:problem_formulation}
% Together, the observations above show that compression effectiveness is highly pair-dependent and varies at the tensor level rather than at the model granularity. 

\phead{Problem Definition.} Minimizing model storage requires jointly choosing which tensors act as bases and which are stored as compressed deltas, a combinatorial optimization shaped by pairwise compressibility estimates that are themselves costly to compute. We formulate this problem below and highlight three key challenges.

%\phead{Optimization Objective and Challenges.}
%At a high level, the goal is to decide which tensors should be stored as \emph{bases} and which ones should be stored as \emph{deltas} to minimize total \emph{storage space cost}.
%Storing a tensor as a base incurs its full size, while storing it as a delta incurs a smaller size but this data reduction depends on which base tensor to perform the delta compression against.
%choosing a good base.
%A base tensor is stored in full.A delta tensor is smaller, but its size depends on which base it is compressed against.
%Storing a tensor as a base requires keeping its full size, whereas storing it as a delta reduces the storage cost but the data reduction achieved by delta compression depends on the specific base tensor it references.

\phead{Optimization Objective.} Storing a tensor as a base incurs its full size; storing it as a compressed delta reduces cost, but the reduction depends on the specific base it references.
Formally, the optimization problem is:
\vskip 1.5em
\[
\tikzmarknode{mbs}{\highlight{green!75!black}{\min_{\text{base selection}}}}
\Big(
\sum \tikzmarknode{cb}{\highlight{red}{cost_{base}}} \;+\; \sum \tikzmarknode{cd}{\highlight{blue}{cost_{delta}}}
\Big).
\]
\begin{tikzpicture}[overlay,remember picture,>=stealth,nodes={align=left,inner ysep=1pt},<-]
    % For "base selection"
    \path (mbs.north) ++ (0,1.2em) node[anchor=south east,color=green!75!black] (mbstext){\textsf{\footnotesize pairing decisions}}; 
    \draw [color=green!75!black](mbs.north) |- ([xshift=-0.3ex,color=purple]mbstext.south west);
    % For "cost_base"
    \path (cb.south) ++ (0,-0.8em) node[anchor=north east,color=red!67] (cbtext){\textsf{\footnotesize base storage cost}};
    \draw [color=red!57](cb.south) |- ([xshift=-0.3ex,color=red]cbtext.south west);
    % For "cost_delta"
    \path (cd.south) ++ (0,-0.8em) node[anchor=north west,xshift=-.3em,align=left,font=\sffamily\footnotesize,color=blue!67] (cdtext){comp. delta\\storage cost};
    \draw [color=blue!57](cd.south) |- ([xshift=-0.3ex,color=blue]cdtext.south east);
\end{tikzpicture}
\vskip 1em 

%This objective closely resembles a \emph{facility location} problem~\cite{expClock2017,Lagrangean4FLP}.
%selecting a base corresponds to \textit{opening} a facility, and assigning a tensor to that base (i.e., storing it as a delta) corresponds to connecting a client to a facility.
%choosing which facility it connects to.
%Opening a facility (selecting a base) incurs a fixed cost equal to the storage size of the base tensor,
%’s byte cost in storage space,
%while assigning a client to a facility incurs a variable cost equal to the delta size when compressed against that base.
%The optimization aims to choose which facilities to open and how to assign clients so that the sum of facility cost and assignment cost is minimized.
\subsection{Optimization Challenges}
\label{sec:optimization}
\label{subsec:challenges}
The optimization objective is an instance of the \emph{facility-location problem}~\cite{expClock2017,Lagrangean4FLP}, which is NP-hard and typically requires heavy integer linear programming (ILP) solvers~\cite{gurobi,cplex,scip}. Solving it at scale is challenging for three reasons:   

% \begin{itemize}[noitemsep,leftmargin=*]
\phead{Challenge \#1 (\clabel{C1}): Estimating pairwise tensor similarity is expensive.}
Determining whether tensor $t_i$ is similar to tensor $t_j$ requires loading both tensors in full---each ranging from millions to billions of parameters---making brute-force pairwise comparisons prohibitively costly at the model-hub scale. 
A natural alternative is approximate nearest-neighbor (ANN) search~\cite{LSH2004,NavigateGraph2019,QueryDrivenGraph2012,smallWorldGraph}, but it suffers from the well-known \emph{curse of dimensionality}~\cite{beyer1999nearest,indyk1998approximate}: as dimensionality grows into the millions, distance concentration renders nearest-neighbor queries no more informative than random selection, and the memory overhead of indexing billion-scale tensors as high-dimensional vectors is itself prohibitive. 
% These limitations motivate {\tensorsketch}, a lightweight fingerprint that can estimate pairwise similarity from compact sketches without accessing full tensor contents (\sref{subsec:tensorsketch}). \tf{add fingerprint discussion}

\phead{Challenge \#2 (\clabel{C2}): Similarity does not imply compressibility.}
Even with an efficient similarity measure, a fundamental gap remains: standard distance metrics (e.g., $L_2$, cosine) operate at the value level and do not capture \emph{bit-level} patterns that govern delta compressibility. Two tensors that are close in $L_2$ space may still produce poorly compressible deltas because their byte-level entropy patterns differ.
% Moreover, even a compression-aware distance only provides a relative ranking of tensor pairs, not the \emph{quantitative reduction ratio} that our optimization formulation requires. 
% The mapping from bit-level similarity to compressibility is non-linear, so a simple threshold or linear scaling is insufficient.

% We bridge this gap with {\tensorpred}, a lightweight regression model that transforms the normalized Hamming distance produced by {\tensorsketch} into a predicted reduction ratio, achieving 0.993 Pearson correlation across millions of tensor pairs (\sref{subsec:compression_ratio_prediction}).

\phead{Challenge \#3 (\clabel{C3}): Selecting bases is combinatorial and dynamic.}
Even with accurate pairwise compressibility estimates, choosing which tensors to serve as bases remains NP-hard~\cite{expClock2017,Lagrangean4FLP}. 
For example, a projection tensor in LLaMA2-7B can choose from over 120,000 possible base tensors. 
% Concretely, selecting a base corresponds to \emph{opening a facility} (incurring its full storage size as a fixed cost), and assigning a tensor to that base corresponds to \emph{connecting a client} (incurring the compressed delta size as a variable cost); the goal is to minimize the total of both costs.
Exact solvers such as ILP~\cite{gurobi} are computationally intractable beyond a few hundred tensors.
Classical approximation algorithms, including primal-dual~\cite{JainV01} and local-search methods~\cite{AryaGKMMP04}, offer theoretical guarantees but still require materializing the full pairwise cost matrix and are designed for static inputs.

Moreover, the tensors stored in real-world model hubs are dynamic: as new models arrive, previously selected bases may become suboptimal, triggering cascading reassignments that require recompressing and re-pairing of affected tensors.
% A practical system must therefore operate \emph{incrementally}---absorbing new tensors without recomputing from scratch---and adapt base selections as the model storage evolves (\sref{subsec:flexsplit}). 

Together, these challenges call for a system that can efficiently estimate pairwise similarity without full-tensor access (C1), predict compressibility from similarity (C2), and select bases at scale in a dynamic, incremental fashion (C3).
% \end{refine}
% \input{sections/overview}
\section{\system Design} 
\label{sec:design}

% \begin{refine}
\subsection{System Overview}
\label{subsubsec:overview_workflow}

\begin{figure}[t]
    \centering
    \includegraphics[width=0.475\textwidth]{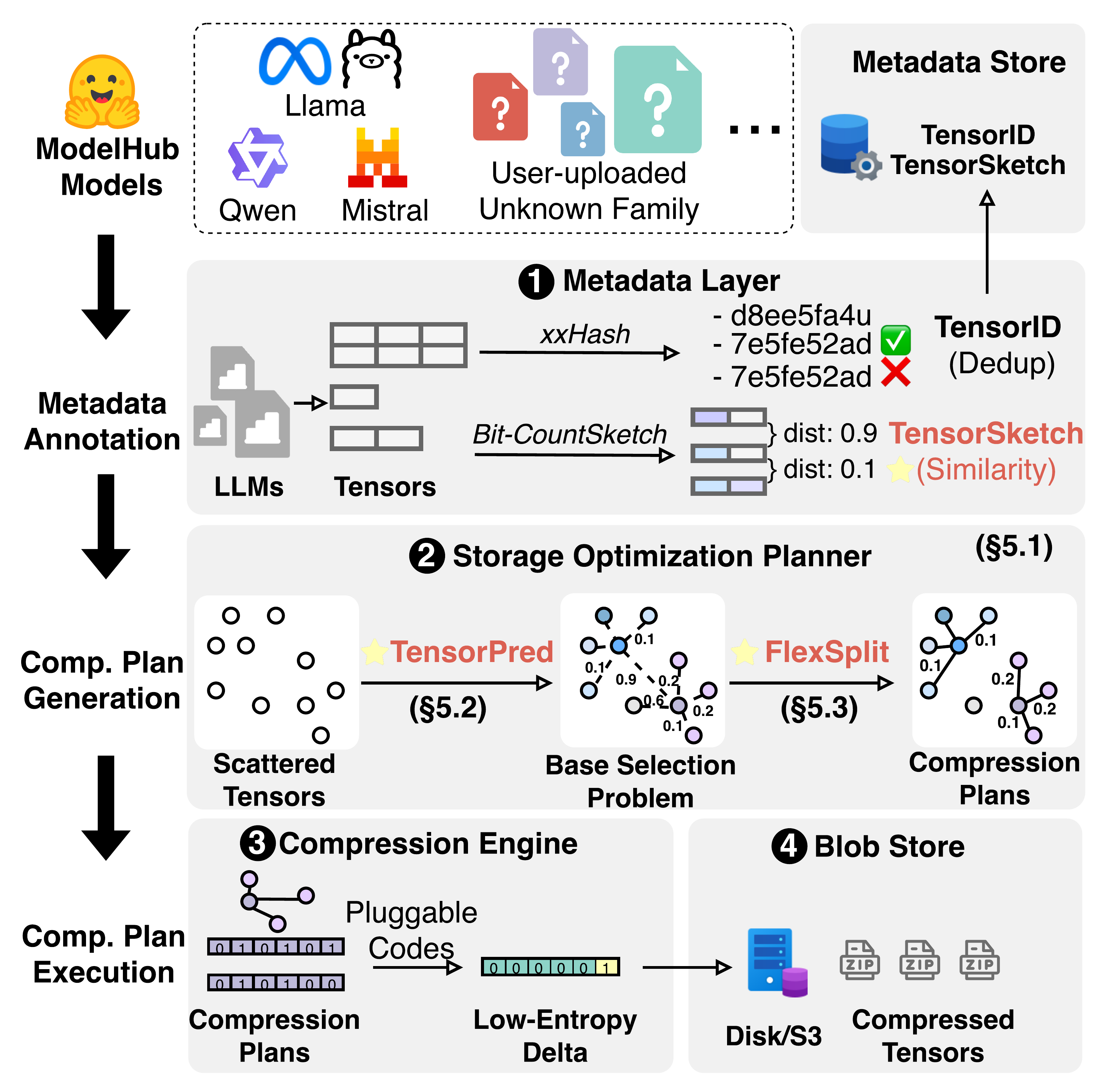}
    \vspace{-15pt}
    \caption{\system architecture and workflow.}
    \label{fig:workflow}
    %\vspace{-1pt}
\end{figure}

\phead{Architecture.} {\system} treats tensors as first-class entities and explicitly reasons about cross-tensor relationships across a model hub, pairing each tensor with a similar base and delta-compressing it to minimize storage footprint.
As shown in \fref{fig:workflow}, \system comprises four components: \circled{1}~a content-aware tensor metadata layer ({\tensorid} and {\tensorsketch}), \circled{2}~a compression planner ({\tensorpred} and {\flexsplit}), \circled{3}~a compression engine ({\tensorx and \fmplus} ), and \circled{4}~a storage layer for tensor blobs on generic backends such as local disk or cloud object stores (e.g., S3~\cite{aws_s3}, R2~\cite{cloudflare_r2}).

\phead{Workflow.} When a batch of models is uploaded, the metadata layer \circled{1} parses each model into individual tensors and assigns two types of metadata: a {\tensorid} content hash for exact deduplication and a compact {\tensorsketch} fingerprint for similarity estimation (\clink{C1}). The compression planner \circled{2} then uses {\tensorpred} to convert fingerprint distances into predicted storage reduction ratios (\clink{C2}), and {\flexsplit} to derive a base-delta clustering plan (\clink{C3}). The compression engine \circled{3} materializes this plan by computing and compressing deltas, and writing blobs to the storage layer \circled{4}.
When a model is requested, the storage layer and compression engine cooperate to fetch bases, apply deltas, and reconstruct full tensors on demand.

\phead{Content-aware Tensor Metadata.}
{\tensorid} is a hash~\cite{xxhash} computed over raw bytes to detect exact duplicates.
{\tensorsketch} is a compact CountSketch-based~\cite{countsketch} fingerprint that captures the bit-level distribution of tensor data, enabling efficient estimation of pairwise normalized Hamming distances (\sref{subsec:tensorsketch}).
% Together, dual metadata provides the foundation for redundancy detection and compression planning.

\phead{Compression Planner.}
The planner determines which tensors should be stored as bases and which should reference others via deltas, thereby achieving a low storage footprint for both bases and deltas. 
It relies on two methods: {\tensorpred} (\sref{subsec:compression_ratio_prediction}), a lightweight regression model that converts {\tensorsketch} distances into predicted reduction ratios; and {\flexsplit} (\sref{subsec:flexsplit}), an incremental clustering algorithm that uses these predictions to select bases and assign tensors.
% without exhaustive pairwise search or heavy ILP solvers.

\phead{Compression Engine.} The compression engine carries out planner-generated compression plans by computing and compressing deltas (\sref{subsec:tensorx}) to generate the final compressed blobs. The engine is fully decoupled from plan generation and can be executed in parallel.

\phead{Storage Layer.}
{\tensorid}, {\tensorsketch}, and model-tensor mappings are stored in a 
%lightweight 
metadata database (PostgreSQL~\cite{postgresql})
%(SQLite in our prototype, PostgreSQL for distributed deployments), 
supporting fast deduplication checks and fingerprint-based similarity queries at ingestion time.
Each tensor blob is stored in a safetensors-compatible format \cite{safetensors} extended with compression flags, unifying compressed and uncompressed tensors under the same format.

\subsection{{\tensorsketch}: Compression-aware Fingerprint}
\label{subsec:tensorsketch}

%old intro:
%The goal of \tensorsketch is to provide a compact summary of each tensor that preserves the information to identify similar tensors: \added{(1) bit distribution
%and (2) tensor layout. These two aspects jointly determine whether two tensors are good candidates for pairing and delta compression.}
%\added{Concretely, \tensorsketch applies a bit-level CountSketch that hashes each element’s bit positions into sketch buckets while incorporating element indices and bit offsets as positional signals, thereby capturing both bit distribution and layout despite CountSketch’s inherent layout obliviousness.}

The goal of {\tensorsketch} is to provide a compact fingerprint for each tensor that captures the bit-level numerical distribution and structural layout---the two aspects that jointly determine whether two tensors are good candidates for delta compression.

%old design:
%\jason{no need to talk about countsketch at beginning}
%Traditional CountSketch operates on floating-point values and preserves a tensor’s \emph{value distribution}. However, bit-level delta compression works at the \emph{bit level}: what matters is not numerical distance per se, but the fraction of bits that differ between two tensors, i.e., the normalized Hamming distance over their raw binary representations. A fingerprint that directly captures per-bit flip statistics is therefore a more faithful proxy for delta compressibility than one based on scalar magnitudes.

\phead{{\tensorsketch} Design.}
Existing fingerprinting approaches fall short for our purpose.
% \tf{content hash - lsh - tensorsketch} Content hashes (e.g., xxHash) detect exact duplicates but are useless for near-duplicate tensors---a single bit flip produces a completely different hash.
Standard CountSketch~\cite{countsketch} operates on floating-point values and preserves \emph{value distribution}, but delta compression works at the \emph{bit level}: what matters is not numerical distance per se, but the fraction of bits that differ between two tensors, i.e., the normalized Hamming distance~\cite{hamming_distance_bell1950} over their raw binary representations.
A fingerprint that directly captures per-bit flip statistics is therefore a more faithful proxy for delta compressibility than one based on scalar magnitudes.
We therefore introduce {\tensorsketch}, a bit-level CountSketch construction that hashes each element’s individual bit positions into sketch buckets, capturing both bit distribution and layout in a compact, fixed-size fingerprint.

\begin{figure}[t]
    \centering
    \includegraphics[width=0.475\textwidth]{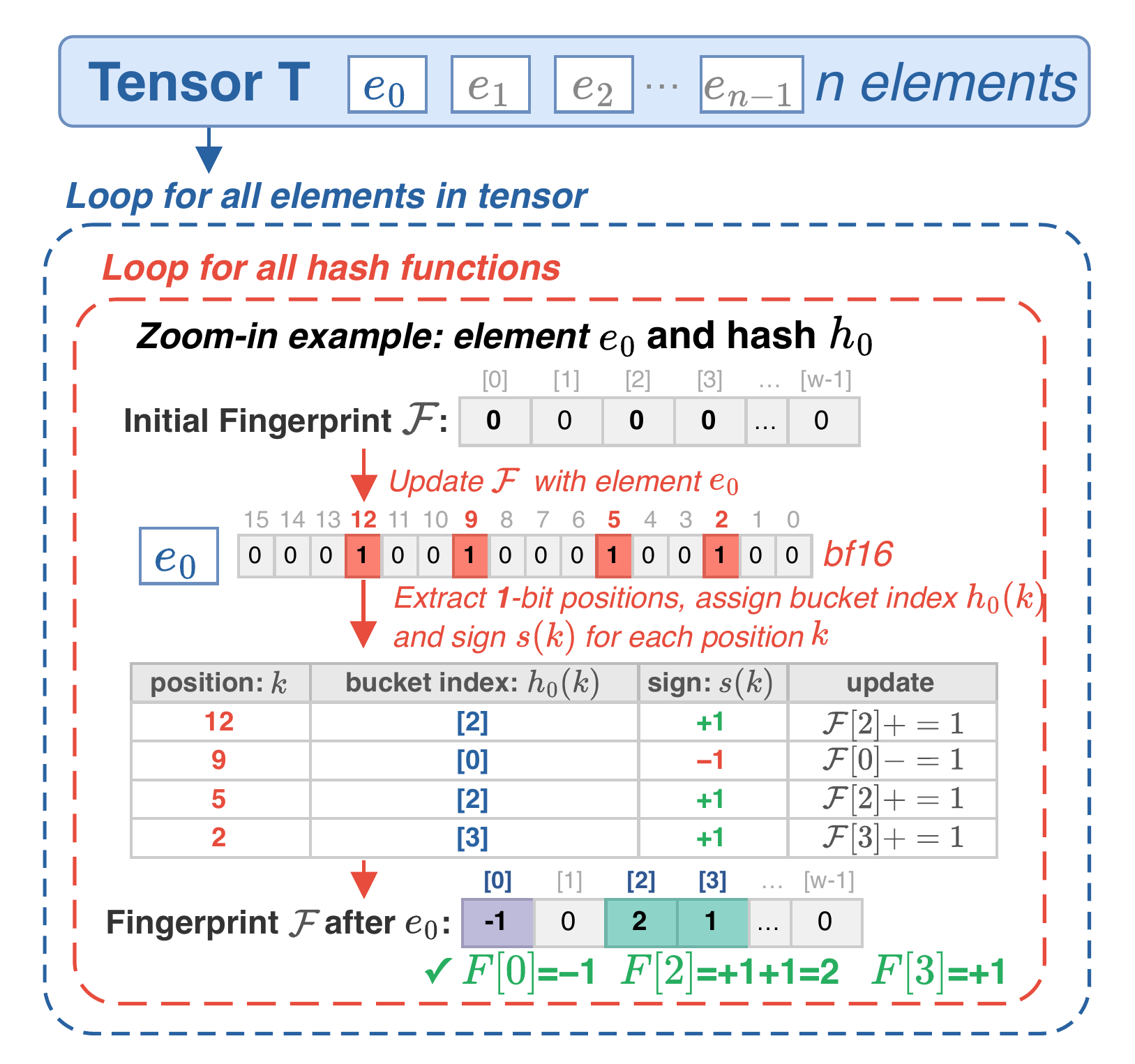}
    \vspace{-15pt}
    \caption{An example of {\tensorsketch} fingerprinting.} 
    \label{fig:tensorsketch}
    %\vspace{-1pt}
\end{figure}

\phead{\tensorsketch Construction.} As shown in \fref{fig:tensorsketch} and \aref{alg:tensorsketch}, \tensorsketch addresses this by treating the tensor's raw bytes as a flat bit sequence of $p \ast n$ bits, grouped into $n$ elements of $p$ bits each (e.g., $p{=}16$ for \texttt{float16}/\texttt{bfloat16}, $p{=}32$ for \texttt{float32}), and building a two-dimensional sketch table $\mathcal{F} \in \mathbb{R}^{d \times w}$ where $d$ is the depth (number of independent hash rows) and $w$ is the width (number of buckets per row). Crucially, each of the $p$ bit positions in every element is treated as an independent binary signal: for bit position $k \in \{0,\dots,p{-}1\}$ of element $i$, a dedicated pair of hash functions $(h_r,\, s_r)$—seeded independently per row $r$—maps the element–bit-position pair $(i, k)$ to a bucket and a sign. This \emph{per-bit hashing} ensures that structural differences at any bit position (exponent, mantissa, or sign) are reflected in the sketch, making \tensorsketch sensitive to the bit-level divergence that drives delta compression efficiency.

\phead{Hamming Distance Estimation.} Given two fingerprints $\mathcal{F}^i$ and $\mathcal{F}^j$, the Hamming distance between the underlying tensors is estimated from the per-row squared $\ell_2$ norm of the difference $\mathcal{F}^i_r - \mathcal{F}^j_r$. Taking the \emph{median} over all $d$ rows provides robustness against outlier hash collisions:

\vskip 1.6em
\begin{equation}
  \mathcal{\hat{H}}(i,j) = \tikzmarknode{med}{\highlight{purple}{\mathrm{median}_{r=1}^{d}}}\; \bigl\|\tikzmarknode{tsi}{\highlight{red}{\mathcal{F}^i_r}} - \mathcal{F}^j_r\bigr\|_2^2.
\end{equation}

\begin{tikzpicture}[overlay,remember picture,>=stealth,nodes={align=left,inner ysep=1pt},<-]
    % For "median"
    \path (med.north) ++ (0,1.5em) node[anchor=south east,color=purple!67] (medtext){\textsf{\footnotesize aggregation for robustness}};
    \draw [color=purple!57](med.north) |- ([xshift=-0.3ex,color=purple]medtext.south west);
    % For "TS^i"
    \path (tsi.south) ++ (0,-1.2em) node[anchor=north west,color=red!67] (tsitext){\textsf{\footnotesize sketch fingerprint}};
    \draw [color=red!57](tsi.south) |- ([xshift=-0.3ex,color=red]tsitext.south east);
\end{tikzpicture}
\vskip 1.6em
\noindent The normalized Hamming distance, $\hat{p}(i,j) = \hat{H}(i,j)\,/\,(pn)$, lies in $[0, 1]$: near zero indicates near-identical bit patterns (highly compressible bit-level delta), while $\approx\!0.5$ indicates random divergence.

% \phead{Fingerprint Generation.}
% TensorSketch extends CountSketch by operating on the bit-level representation of tensor elements.

\begin{algorithm}[t]
\caption{\tensorsketch Fingerprinting} 
\label{alg:tensorsketch}
\KwIn{\parbox[t]{0.85\linewidth}{%
  \textit{Params:} sketch depth $d$; sketch width $w$\\[2pt]
  \textit{Data:} Tensor $t$ with $(e_0,\dots,e_{n-1})$, each $p$ bits\\[2pt]
  \textit{Hash functions:} hash families $\{h_r\}_{r=1}^{d}$ with $h_r\!: [n]\!\times\![p]\!\to\![w]$ and $\{s_r\}_{r=1}^{d}$ with $s_r\!: [n]\!\times\![p]\!\to\!\{-1,\!+1\}$%
}}
\KwOut{\tensorsketch $\mathcal{F} \in \mathbb{R}^{d \times w}$.}
% $(v_1,\dots,v_n) \leftarrow$ group bits of $t$ into $p$-bit unsigned elements\;
Initialize $\mathcal{F}[0..d-1][1..w] \leftarrow 0$\;
\For{$i \leftarrow 0$ \KwTo $n-1$}{
    \If{$e_i = 0$}{\textbf{skip}\tcp*{all bits zero}}
    \For{$r \leftarrow 0$ \KwTo $d-1$}{
        \For{each bit position $k$ where $e_i[k] = 1$}{
            $j \leftarrow h_r(i,k)$ \tcp*{bucket index}
            $\sigma \leftarrow s_r(i,k)$ \tcp*{sign}
            $\mathcal{F}[r][j] \leftarrow \mathcal{F}[r][j] + \sigma$\;
        }
    }
}
\Return $\mathcal{F}$\;
\end{algorithm}

% \end{refine}

% \begin{refine}

%\subsection{Reduction Ratio Prediction Model}
\subsection{{\tensorpred}: Reduction Ratio Prediction}
\label{subsec:compression_ratio_prediction}

%old intro:
% \added{Estimating the compression benefit of pairing two tensors is critical for guiding {\salgo}’s assignment and splitting decisions. In particular, the planner must efficiently approximate the storage cost of compressing one tensor as a delta of another, without materializing the actual delta. To avoid the prohibitive computation and I/O cost of explicit delta evaluation (C2), \system learns a lightweight prediction model that estimates reduction ratios using only the normalized Hamming distance between {\tensorsketch} fingerprints.}

{\tensorsketch} estimates pairwise similarity, but the optimization formulation in \sref{sec:problem_definition} requires \emph{quantitative reduction ratios} to evaluate each base-delta assignment (\clink{C2}).
{\tensorpred} bridges this gap with a lightweight regression model that converts {\tensorsketch} distances into predicted reduction ratios without materializing any actual delta.

\phead{Key Observation.}
Under bit-level delta compression (detailed in \sref{subsec:tensorx}), the fraction of differing bits between two tensors reflects the entropy of the resulting residuals and hence the achievable reduction ratio. 
The normalized Hamming distance $\hat{p}$ estimated by {\tensorsketch} correlates with this fraction. 
However, the relationship between $\hat{p}$ and reduction ratio is non-linear: at small $\hat{p}$, a marginal increase in Hamming distance disproportionately raises the fraction of large-magnitude residuals, causing a sharp drop in reduction; at larger $\hat{p}$, residuals are already broadly distributed and additional divergence has diminishing effect---reduction ratio approaches 0 (no gain). Binary entropy $\mathcal{S}(\hat{p}) = -\hat{p}\log_2\hat{p} - (1{-}\hat{p})\log_2(1{-}\hat{p})$, scaled as $\tau = 8\,\mathcal{S}(\hat{p})$, captures the per-element bit uncertainty a compressor must resolve and naturally models this non-linear regime change.

% \rev{Neither term alone suffices: $\hat{p}$ is linear, while $\tau$ is symmetric about $\hat{p}{=}0.5$ and cannot tell $\hat{p}{=}0.1$ from $\hat{p}{=}0.9$. Their product keeps the entropic bend without that ambiguity.}

\phead{Model Formulation.} Let $\hat{p}$ denote the normalized Hamming distance estimated by the {\tensorsketch} distance estimator (\sref{subsec:tensorsketch}). We define the following auxiliary feature:  
\vspace{-4pt}
\begin{equation}
  \tau = 8\,\mathcal{S}(\hat{p}),
\vspace{-4pt}
\end{equation}
where $\mathcal{S}(\cdot)$ is the binary entropy function and the factor of 8 normalizes $\tau$ to the per-byte bit uncertainty. The predicted reduction ratio is:
\vskip 1.6em
\begin{equation}
\mathcal{R}(\hat{p}) = \tikzmarknode{pterm}{\highlight{red}{\alpha\,\hat{p}}} + \tikzmarknode{tterm}{\highlight{blue}{\beta\,\tau}} + \tikzmarknode{cross}{\highlight{purple}{\gamma\,(\hat{p} \cdot \tau)}} + \tikzmarknode{bias}{\highlight{OliveGreen}{\epsilon}},
\end{equation}
\begin{tikzpicture}[overlay,remember picture,>=stealth,nodes={align=left,inner ysep=1pt},<-]
    % For "alpha p"
    \path (pterm.north) ++ (0,1.5em) node[anchor=south east,color=red!67] (ptermtext){\textsf{\footnotesize bit divergence}};
    \draw [color=red!57](pterm.north) |- ([xshift=-0.3ex,color=red]ptermtext.south west);
    % For "beta t"
    \path (tterm.north) ++ (0,1.5em) node[anchor=south west,color=blue!67] (ttermtext){\textsf{\footnotesize entropy cost}};
    \draw [color=blue!57](tterm.north) |- ([xshift=-0.3ex,color=blue]ttermtext.south east);
    % For "gamma (p * t)"
    \path (cross.south) ++ (0,-1.2em) node[anchor=north east,color=purple!67] (crosstext){\textsf{\footnotesize nonlinear correction}};
    \draw [color=purple!57](cross.south) |- ([xshift=-0.3ex,color=purple]crosstext.south west);
    % For "delta"
    \path (bias.south) ++ (0,-1.2em) node[anchor=north west,color=OliveGreen!67] (biastext){\textsf{\footnotesize bias}};
    \draw [color=OliveGreen!57](bias.south) |- ([xshift=-0.3ex,color=OliveGreen]biastext.south east);
\end{tikzpicture}
\vskip 1.6em 
where $(\alpha, \beta, \gamma, \epsilon)$ are regression coefficients fitted offline on a corpus of real tensor pairs with their measured reduction ratios, and the result is clipped to $[0, 1]$. The three terms provide complementary signals: $\hat{p}$ captures raw bit divergence, $\tau$ captures its information-theoretic cost, and $\hat{p} \cdot \tau$ models the nonlinear coupling between divergence magnitude and entropic uncertainty (see \fref{fig:pred_model_val}).

% \end{refine}

\subsection{\salgo: Adaptive Tensor Clustering and Cluster Splitting}
\label{subsec:flexsplit}

%old intro:
%While \system formulates the large-scale tensor compression optimization
%as a facility-location problem (\sref{sec:problem_definition}), solving this problem at scale is \added{challenging}:
%adding a new model can change which bases are optimal, requiring re-evaluating delta sizes across many tensors and potentially re-running the global optimization.
%To support continuous ingestion while preserving most of the compression benefit, we design {\salgo} as a highly efficient, lightweight online heuristic algorithm
%that operates entirely in the fingerprint space without ever reading full tensors.
%\salgo follows a two-phase strategy:
%starting off, it greedily assigns each arriving tensor to the cheapest existing base using predicted delta costs; it then periodically refines these assignments by splitting overgrown clusters into a small number of coherent centers.

% As discussed in \clink{C3}, base selection is NP-hard, ILP solvers~\cite{gurobi,cplex,scip} do not scale, and classical heuristics~\cite{JainV01,AryaGKMMP04} assume static inputs that cannot adapt to continuously arriving models.
Finding the best base is an NP-hard problem and ILP solvers are too slow for online systems. We design 
{\salgo}, a two-phase online algorithm that operates entirely on {\tensorsketch} fingerprints and {\tensorpred} estimates. First, it greedily assigns each arriving tensor to the cheapest existing base, then periodically splits overgrown clusters.

\if 0

\phead{Notation.} Let $s_t$ denote the raw byte size of tensor $t$, and let $\mathcal{F}^t$ denote its {\tensorsketch} fingerprint. We write $\mathcal{R}(t, b)$ for the \emph{predicted reduction ratio} of tensor $t$ against base $b$, given by {\tensorpred}.
% evaluated at the {\tensorsketch} distance $\hat{p}(t,b)$ 
The predicted storage cost of encoding $t$ as a delta against $b$ is therefore $(1 - \mathcal{R}(t, b)) \cdot s_t$.

\fi

\begin{figure}[t]
    \centering
    \includegraphics[width=0.475\textwidth]{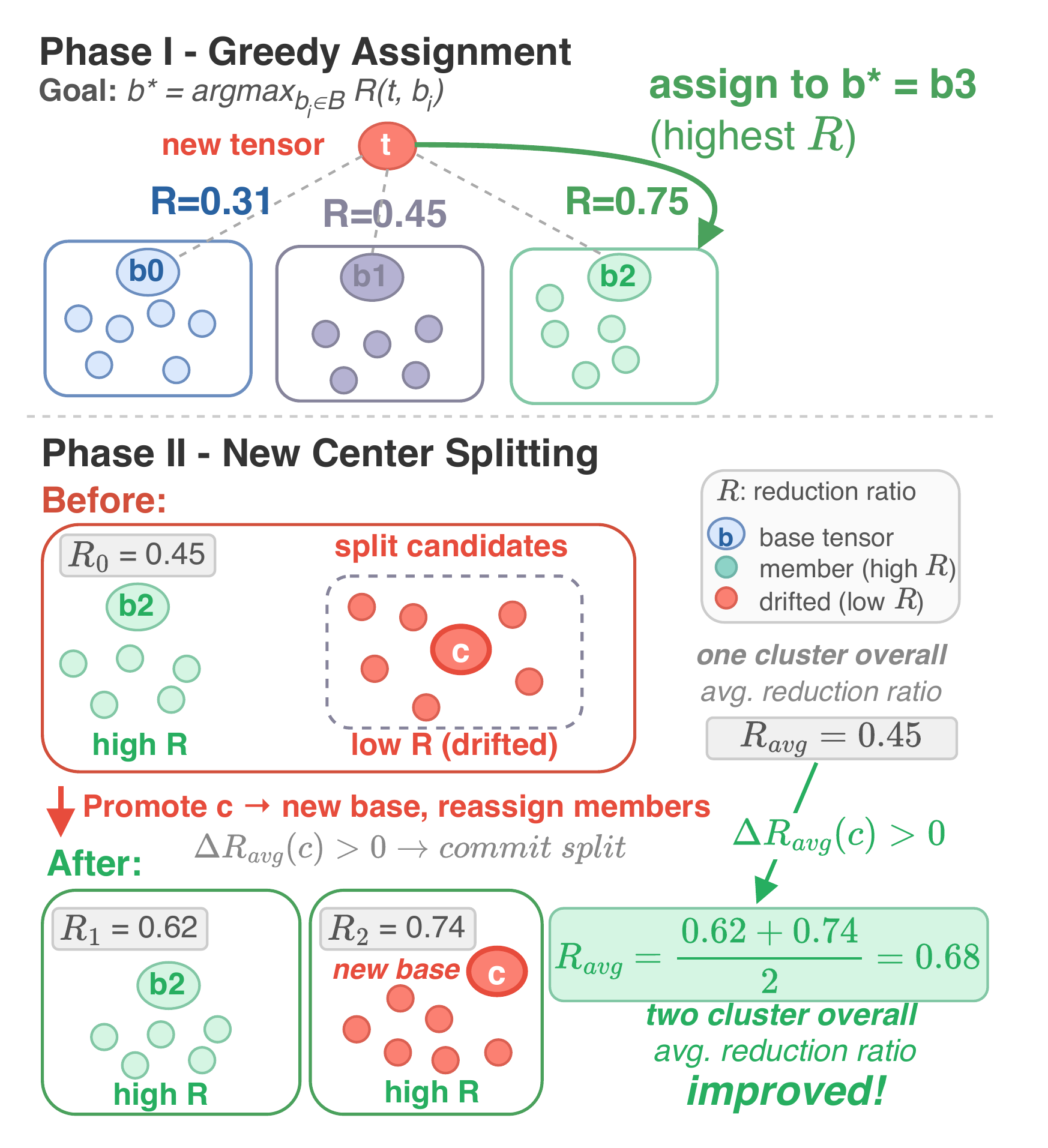}
    \vspace{-15pt}
    \caption{Example workflow of {\flexsplit}.} 
    %\added{$R$ means reduction ratio.}} 
    \label{fig:flexsplit}
    %\vspace{-1pt}
\end{figure}

\phead{Phase I: Greedy Incremental Assignment.}
As new models are uploaded to the model hub,
\system processes each model's tensors in turn.
The first tensor with unseen shape and name naturally seeds a new cluster as its initial base. For every subsequent tensor, \salgo determines which existing base yields the highest predicted reduction ratio and assigns the tensor to that base's cluster.
As illustrated in \fref{fig:flexsplit}~(top), a new tensor $t$ is evaluated against three existing bases $b_0$, $b_1$, $b_2$ with predicted reduction ratios $0.31$, $0.45$, and $0.75$, respectively; \salgo assigns $t$ to $b_2$ because it offers the highest predicted reduction.
Since all comparisons operate on compact {\tensorsketch} fingerprints rather than raw tensor bytes, each assignment requires at most $O(k)$ distance evaluations over the current base set $\mathcal{B}$, where $k = |\mathcal{B}|$ is typically small. With approximate nearest-neighbor indexing, this cost further reduces to $O(\log k)$ per tensor.

\phead{Improvement Opportunities from New Arrivals .}
Greedy assignment connects each tensor to the best base \emph{available at the time of ingestion}. As new models arrive, however, they introduce tensors that may serve as better centers for existing members. In other words, the new arriving tensors improve the solution space that the current static assignments do not reflect. \fref{fig:flexsplit}~(Phase II, ``Before'')  
illustrates this: the cluster around $b_2$ mixes well-matched members (green, high $\mathcal{R}$) with poorly-matched ones (red, low $\mathcal{R}$) that would compress
better against a newly available center, dragging the cluster-wide average $\overline{\mathcal{R}}$ down to $0.45$.                                                     
To capture these improvement opportunities without rerunning global optimization, \salgo periodically revisits each cluster and asks whether promoting an existing member
to a new base would reduce the projected storage cost.

\phead{Phase II: Multi-Center Splitting.} Phase~II revisits each cluster and asks whether promoting an existing member to a new base would improve the cluster-wide reduction ratio, defined as the total bytes saved by delta-encoding non-base tensors against their best base, divided by the total raw size of all members. Bases themselves are stored in full and contribute zero reduction. Initially each cluster has a single base, so the initial ratio serves as the baseline.

\salgo solves this greedily. Starting from the original base, it first identifies \emph{split candidates}: tensors whose predicted reduction ratio falls significantly below the cluster average reduction ratio, indicating poor alignment with the current center. For each candidate, it evaluates the gain of promoting that tensor to a new base. Promoting a candidate trades off its own delta savings (now stored in full) against higher reduction ratios for nearby tensors that better align with it. A split is committed only when the split benefit is positive. \salgo selects the candidate with the largest gain, promotes it to a new base, and reassigns all members whose reduction ratio improves under this new center.
As shown in \fref{fig:flexsplit}~(bottom, ``After''), promoting candidate $c$ to a new base partitions the original cluster into two coherent sub-clusters---both now exhibit high $\mathcal{R}$ for all members, lifting the cluster-wide average from $\overline{\mathcal{R}} = 0.45$ to $0.68$.

\rev{Splitting is \emph{cluster-local}: the promoted base comes from within the cluster and only its members are re-scored. Unlike an incremental vector index, where a split must reassign boundary vectors in \emph{neighboring} partitions to preserve recall~\cite{spfresh_sosp23,quake_osdi25}, the storage-cost objective decomposes per cluster, so no cascading refinement follows.}

\vspace{-5pt}
\section{Implementation}
\label{sec:impl}
\label{sec:implementation}

{\system} is implemented in roughly 12K lines of Python and 3K lines of optimized C/Rust, and integrates directly with Hugging Face \texttt{safetensors} model formats. This section describes the two main implementation aspects: the 
metadata engine 
%storage format 
that organizes the metadata of base and delta tensors 
%for efficient retrieval 
(\sref{subsec:impl_format}), 
and the SIMD/AVX2-accelerated operators that make ingestion and decompression fast (\sref{subsec:impl_operators}).

%\subsection{Storage Format}
\subsection{Metadata Engine} 
\label{subsec:impl_format}

%\phead{Metadata Store.}
\phead{Metadata Schema.} {\system} maintains per-tensor metadata in PostgreSQL for distributed deployments. 
Each tensor metadata entry stores: 

\begin{center}
\texttt{(model\_id, tensor\_name, tensor\_id, tensor\_sketch, dtype, shape)}
\end{center}

\noindent As the original safetensors format already includes \texttt{dtype} and \texttt{shape} in its header, {\system} extracts them at zero cost. These fields allow the planner to filter out structurally incompatible candidates before computing any fingerprint distances, reducing the search space for valid base-delta pairs.

\rev{{\tensorid} is a 16\,B xxHash128~\cite{xxhash} digest over the tensor's raw bytes, used both as the deduplication key and as the content address that a decoded tensor is verified against at retrieval time. {\tensorsketch} uses $d{=}2$ rows of $w{=}1024$ 32-bit counters, i.e., 8\,KB per tensor (\tref{tab:metadata}).}

\phead{Search Index.} 
To efficiently retrieve the most similar base for a given tensor, {\system} builds an HNSW~\cite{HNSW2018} index \emph{over {\tensorsketch} fingerprints}. At ingestion time, each new tensor's sketch is inserted into the index; at query time, the planner performs an approximate nearest-neighbor lookup to identify candidate bases in $O(\log n)$ time, avoiding exhaustive pairwise comparisons across the full tensor pool.

\begin{revblock}
\phead{Deletions.}
Deletion is deferred: {\system} tombstones the tensor and reclaims its bytes in a later offline maintenance transaction. If the tensor is a base with live dependents, that transaction promotes another member of its cluster to base and recompresses the dependents against it, swapping the metadata pointer atomically once the new blob is written. The rewrite stays within that one cluster.
\end{revblock}

% \noindent The \textbf{tensor\_id} (xxHash64 over raw bytes) enables fast exact tensor-level deduplication. 
% \noindent The \textbf{tensor\_sig} field is a compact tuple of dtype, shape, and element count that lets the planner filter out incompatible candidates before computing any fingerprint distances, substantially reducing the search space for valid base-delta pairs.

% {\tensorsketch} vectors are small (8\,KB per tensor), so all similarity search and cluster operations remain cache-resident.

\if 0
\phead{Delta Storage Layout.}
Once the planner assigns each tensor to a base, {\system} materializes the compressed delta via {\tensorx} (\cref{subsec:impl_operators}) and stores it alongside base tensors. Each delta record is self-contained---carrying the base tensor reference, element count, and the compressed chunk payloads---so that retrieval requires only a single metadata lookup followed by one sequential read of the base and one of the delta. \yuec{this is confusing: if one base maps to many deltas, how can reading of one base-delta pair be sequential -- on disk, this is supposed to be stored at separate disk sectors, unless a base is mapped to only one delta}
\fi 

\subsection{High-Performance Codec}
\label{subsec:impl_operators}

All compute-intensive operations in the ingestion and retrieval paths are implemented as SIMD-accelerated kernels in C and Rust, enabling {\system} to saturate memory bandwidth on modern hardware. 
% We highlight three key operators and summarize their throughput in \tref{tab:operator_throughput}.

\phead{\tensorid.}
Tensor deduplication hashes each tensor's raw bytes using \rev{xxHash128} with AVX2 vectorization.

\phead{{\tensorsketch}.}
The {\tensorsketch} kernel (\sref{subsec:tensorsketch}) is implemented in Rust as a fused single-pass streaming operator over zero-copy \texttt{mmap} buffers. Per-bit hash and sign lookups are SIMD-parallel, and sketch accumulation requires no intermediate buffers.

\phead{Pluggable Codec Engine.}
\label{subsec:tensorx}
{\system}'s compression engine exposes a pluggable codec interface, decoupling planning from compression so that any lossless compressor can be used.
We implement two codecs targeting different tradeoffs:
\begin{itemize}[noitemsep,leftmargin=*]
\item \textbf{Throughput-oriented {\tensorx}} extends BitX~\cite{zipllm_nsdi26} with chunk-level parallelism and byte-plane separation.
Chunks are compressed independently in parallel; byte grouping (e.g., first byte of different BF16 floats compressed together) is used to exploit the near-zero skew of fine-tuned weight deltas for better entropy coding.
All kernels are SIMD/AVX2-accelerated.
\item \textbf{Reduction-oriented FM++} is a lossless codec for fine-tuned models, extended from FM-Delta~\cite{ning2024fm}. 
%floating-point delta scheme originally designed for scientific data.
We add native \texttt{BF16} support and chunk-level parallel execution, improving throughput from $\sim$100\,MB/s to 9,821\,MB/s (a $\sim$98$\times$ speedup) while retaining its higher reduction ratio over \tensorx.  
\end{itemize} 

\if 0

\begin{table}[t]
\small
\centering
\caption{Operator throughput for in-memory data on \texttt{c6a.48xlarge} (192 threads).} 
\vspace{-8pt}
\label{tab:operator_throughput}
\scalebox{0.9}{
\begin{tabular}{cccc}
\toprule
\textbf{TensorID} & \textbf{\tensorsketch} & \textbf{{\tensorx} encode} & \textbf{{\tensorx} decode} \\
\midrule
67.5\,GB/s & 27.7\,GB/s & 22.9\,GB/s & 28.4\,GB/s \\
\bottomrule
\end{tabular}
}
\vspace{-5pt}
\end{table}

\fi 
\section{Evaluation}
\label{evaluation}

\subsection{Experimental Setup}
\label{subsec:eval_setup} 

% \begin{wraptable}{r}{0.51\columnwidth}
% \centering
%     \caption{Model statistics summary.}
%     \label{tab:model_stats}
%     \scalebox{0.85}{
%         \begin{tabular}{l r}
%             \toprule
%             \textbf{Metric} & \textbf{Value} \\
%             \midrule
%             Model count & \num{3048} \\
%             Total size & \num{43.19}~TB \\
%             Size after file dedup & \num{41.8}~TB \\
%             \bottomrule
%         \end{tabular}
%     } % end of scalebox
% \end{wraptable}

\begin{table}[t]
\centering
\caption{Evaluation dataset. \rev{All reduction ratios are measured against the 40.11\,TB size deduplication based on \tensorid.}}
\label{tab:model_stats}
\small

\begin{tabular}{ccccc}
\toprule
& \multicolumn{2}{c}{\rev{\textbf{Before dedup.}}} & \multicolumn{2}{c}{\rev{\textbf{After dedup.}}} \\
\cmidrule(lr){2-3} \cmidrule(lr){4-5}
\textbf{Models} & \textbf{Tensors} & \rev{\textbf{Size}} & \textbf{Tensors} & \rev{\textbf{Size}} \\
\midrule
2,890 & 961,509 & \rev{44.86~TB} & 741,689 & 40.11~TB \\
\bottomrule
\end{tabular}
\end{table}

\begin{figure*}[t]
    \centering
    \includegraphics[width=0.995\textwidth]{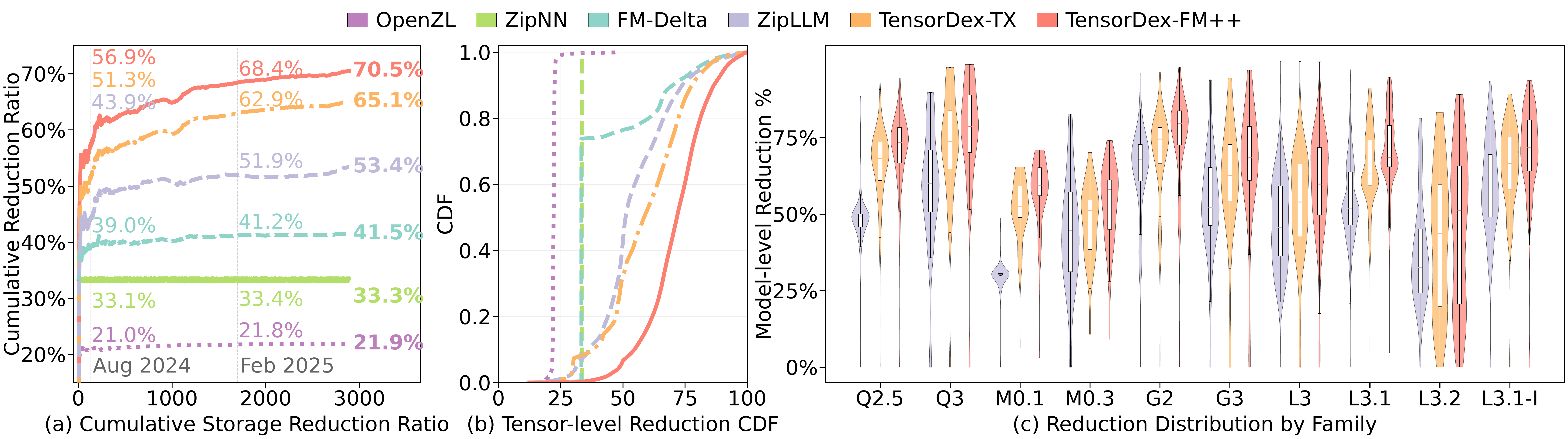}
    \caption{({\bf a}) Cumulative data reduction ratio as models are ingested into the ZipLLM-Trace corpus (ordered by creation time). 
    ({\bf b}) Per-tensor data reduction ratio CDF. 
    ({\bf c}) Per-model reduction ratio distributions by model family (Q: Qwen, M: Mistral, L: Llama, G: Gemma, I: Instruct). 
    TensorDex-FM++ consistently achieves the highest median reduction across all ten families, followed by TensorDex-TX.}
    %TensorDex-FM reaches 70.5\% reduction and TensorDex-TX reaches 65.1\%, both sustaining growth throughout ingestion.
    %ZipLLM plateaus at 53.4\%, while FM-Delta (37.1\%), ZipNN (33.3\%), and OpenZL (21.9\%) remain relatively flat.}
    \label{fig:reduction_global_zipllm_trace}
\end{figure*}

\noindent\textbf{Dataset.} We evaluate {\system} by reconstructing the random-sampled model trace used in ZipLLM~\cite{zipllm_nsdi26}. The details are shown in Table~\ref{tab:model_stats}. 
% After removing 156 entries whose authors have deleted or deprecated the associated models, the resulting dataset contains 2,892 models totaling 40.11\,TB (
It spans a broad range of architectures: 968 models derived from \texttt{Qwen2.5}~\cite{yang2024qwen2}, 151 from \texttt{Qwen3}~\cite{yang2025qwen3}, 139 from \texttt{Mistral}~\cite{jiang2023mistral7b}, 114 from \texttt{Llama-3}~\cite{meta2024llama3}, 1,431 from \texttt{Llama-3.1}~\cite{meta2024llama3p1}, 47 from \texttt{Llama-3.2}~\cite{meta2024llama3-2}, 135 from \texttt{Gemma-2}~\cite{gemma2024}, and 63 from \texttt{Gemma-3}~\cite{team2025gemma}.
\rev{The trace is sampled randomly \emph{within} these representative fine-tuned families rather than across the entire hub. LoRA adapters~\cite{hu2022lora} and quantized checkpoints are tensor sets as well and need no pipeline change beyond a shared \texttt{dtype} between delta and base, but full-weight fine-tunes and checkpoints dominate hub storage (99.1\%, \fref{fig:modelhub_growth}) and are our focus here.}

\noindent \textbf{{\system} Configurations.}
{\system}'s compression engine exposes a pluggable codec interface, allowing different delta compressors to be swapped without changing the planning pipeline. We evaluate two configurations:
\begin{itemize}[noitemsep,leftmargin=*]
    \item \textbf{TensorDex-TX} ({\flexsplit} + {\tensorx}) for optimized compression/decompression throughput. 
    %{\tensorx} extends \bitx with chunk-level parallelism and byte-plane separation, delivering the highest ingestion and retrieval speed.
    \item \textbf{TensorDex-FM++} ({\flexsplit} + FM++) for optimized reduction ratio. 
    %We enhance the original FM-Delta~\cite{} with \texttt{BF16} support and parallel execution, improving its throughput from $\sim$100\,MB/s to over 9,800\,MB/s while retaining its superior compression effectiveness.
\end{itemize}

%\noindent \textbf{Baselines.}
%We evaluate {\system} against baselines spanning \added{three problem dimensions: similarity search, compression planning, and delta codec}.
%four dimensions of the problem: similarity search, compression planning, delta codec, and optimal solving.

\phead{Compression Planner Baselines.}
To evaluate the end-to-end storage reduction achieved by {\system}'s planning pipeline ({\tensorpred} + {\flexsplit}), we compare against production-grade systems:
\begin{itemize}[noitemsep,leftmargin=*]

    \item \textbf{ZipLLM~\cite{zipllm_nsdi26}:} the state-of-the-art model-aware lossless storage reduction system. It clusters models using either (i) the \texttt{base\_model} field
    %model-family metadata 
    from Hugging Face model cards (when available) or (ii) a bit-distance metric~\cite{zipllm_nsdi26} that measures the number of differing bits between the target model and candidate bases,  
    %model-family metadata from model cards 
    and then applies XOR-based delta compression within each cluster. 
    %It falls back to ZipNN~\cite{hershcovitch2024zipnn} when metadata is unavailable. 

    \item \textbf{Default:} since FM-Delta~\cite{ning2024fm} lacks a planning component, we enhance it with Hugging Face \texttt{base\_model} metadata to group models by family and apply delta compression within each group. To reflect real-world metadata availability, we randomly sample 25.8\% of models in the ZipLLM trace to carry metadata; the remaining models fall back to ZipNN~\cite{hershcovitch2024zipnn} for standalone compression.

\end{itemize}

\phead{Compression Codec Baselines.}
To isolate the effect of the delta codec, we compare {\tensorx} and FM-Delta against alternative compressors applied to the same base-delta pairs:
\begin{itemize}[noitemsep,leftmargin=*]
    \item \textbf{BitX~\cite{zipllm_nsdi26}:} XOR-based delta compression that serves as the foundation for {\tensorx}.
    
    \item \textbf{FM-Delta~\cite{ning2024fm}} vanilla: 
    %a floating-point delta compression scheme designed for scientific data. The original implementation 
    which lacks \texttt{BF16} support and runs single-threaded at $\sim$100\,MB/s. 
    %; our enhanced version adds \texttt{BF16} and parallel execution.} 
    
    \item \textbf{ZipNN~\cite{hershcovitch2024zipnn}:} groups floating-point numbers by byte component for single-tensor compression. It does not perform cross-tensor deduplication or delta compression; we additionally enable tensor-level deduplication ({\tensordedup}) for a fair comparison.
    
    \item \textbf{zstd~\cite{zstd}:} a general-purpose lossless compressor widely used in storage systems, applied to raw deltas.
    \item \textbf{OpenZL~\cite{openzl_arxiv25}:} a high-throughput lossless compression library that aggregates multiple compression algorithms.
\end{itemize}

% Note that we did not compare with FM-Delta~\cite{ning2024fm} and {\textsc{Elf}}~\cite{su2024everything}, because FM-Delta does not support \texttt{BF16}, which is the most popular data type for LLMs, and \textsc{Elf} is lossy. 

%At the component level, we compare the baseline compression algorithms and zstd. To isolate the compression benefits attributable solely to each method, we apply our TensorID-based \tensordedup{} uniformly across all baselines, and perform all subsequent compression experiments on this deduplicated dataset. This design ensures that observed differences in compression ratio stem from the algorithms themselves rather than from variability in redundant tensor removal.

%\noindent\textbf{Implementation.} 
%We implemented {\system} in Python and Rust. In total, our implementation comprises over 6,000 lines of code. For ZipLLM, ZipNN, we use the open-source repos from the authors. 

% \noindent\textbf{Metrics.} 
% We evaluate {\system} using multiple metrics. 
% \begin{itemize}[noitemsep,leftmargin=*]
%     \item{\bf Data reduction ratio} (DRR) calculates the data size reduced by deduplication and/or compression over the original data size. A higher reduction ratio is better. 

%     \item{\bf Throughput} measures the speed of deduplication, compression, and decompression of different systems. 

%     \item{\bf Scalability} measures how the storage system scales with the number of models. It primarily concerns solving time for tensor compression plan generation.
% \end{itemize}

\begin{table}[t]
\centering
\caption{Tensor metadata statistics and overhead.}
\label{tab:metadata}
\footnotesize
\begin{tabular}{lccc}
\toprule
\textbf{Method} & \textbf{\#Tensors} & \textbf{Per tensor} & \textbf{Total size} \\
\midrule
TensorID (xxHash128) & 741,689 & 16~B & 11.97~MB \\
TensorSketch ($d{=}2$, $w{=}1024$) & 741,689 & 8~KB & 6.08~GB \\
\midrule
Metadata overhead & \multicolumn{3}{c}{0.015\%} \\
\bottomrule
\end{tabular}

\end{table}

\noindent\textbf{Setup.} 
We conduct our experiments on an Amazon EC2~\cite{aws_ec2} \texttt{c6a.48xlarge} instance, equipped with a 96-core (192-thread) AMD EPYC 7R13 processor and 384\,GB of DRAM.
\rev{Reduction ratios are computed from stored byte counts; the throughputs in \tref{tab:compression_decompression_throughput} keep all data in DRAM; and \tref{tab:storage_retrieval} uses two real backends---Amazon S3~\cite{aws_s3} in the same region, and an NVMe-attached EBS gp3 volume provisioned to 2,000\,MiB/s, read cold and warm.}
Table~\ref{tab:metadata} quantifies {\system}'s metadata overhead.
%All models and associated data are stored in an EBS SSD volume.\yuec{not S3?} 

% \begin{figure}[t]
%     \centering
%     \begin{subfigure}[t]{0.225\textwidth}
%         \centering
%         \includegraphics[width=\linewidth]{figures/evaluation/data_reduction_comparison.pdf}
%         \caption{ZipLLM-Trace.}
%         \label{fig:e2e_eval_zipllm}
%     \end{subfigure}
%     \hfill
%     \begin{subfigure}[t]{0.225\textwidth}
%         \centering
%         \includegraphics[width=\linewidth]{figures/evaluation/data_reduction_comparison_new_trace.pdf}
%         \caption{HF-Random.}
%         \label{fig:e2e_eval_hf}
%     \end{subfigure}
%
%     \caption{End-to-end compression evaluation: (a) ZipLLM-Trace, and (b) HF-Random.}
%     \label{fig:e2e_eval}
% \end{figure}

% \begin{figure}[t]
%     \centering
%     \includegraphics[width=0.475\textwidth]{figures/evaluation/data_reduction_comparison_by_family.pdf}
%     \caption{Data reduction ratio trends by model families. }
%     \label{fig:ddr_breakdown_trend}
% \end{figure}

% \begin{figure}[t]
%     \centering
%     \includegraphics[width=0.475\textwidth]{figures/evaluation/reduction_global.pdf}
%     \caption{Cumulative data reduction ratio as models are ingested into the ZipLLM-Trace corpus (ordered by creation time).
%     TensorDex-FM reaches 70.5\% reduction and TensorDex-TX reaches 65.1\%, both sustaining growth throughout ingestion.
%     ZipLLM plateaus at 53.4\%, while FM-Delta (37.1\%), ZipNN (33.3\%), and OpenZL (21.9\%) remain relatively flat.}
%     \label{fig:reduction_global_zipllm_trace}
% \end{figure}

\subsection{End-to-end Comparison}
\label{sec:e2e_eval}

\subsubsection{Data Reduction Ratio}

To assess the end-to-end effectiveness of {\system}, we run both TensorDex-TX and TensorDex-FM++ over the ZipLLM-Trace dataset.
To mimic real-world usage on model hubs such as Hugging Face---where models arrive continuously---we incrementally ingest models 
in creation order and record the cumulative data reduction ratio. \fref{fig:reduction_global_zipllm_trace}(a) reports how storage savings evolve as the corpus grows. 
%and \fref{fig:reduction_global_zipllm_trace}(b) plots end-to-end reduction. 

Both {\system} variants consistently outperform all baselines and continue improving as more models are ingested, with no signs of saturation.
TensorDex-FM++ achieves the highest reduction ratio (70.5\%) owing to FM-Delta’s stronger compression, while TensorDex-TX reaches 65.1\% with substantially higher throughput (\sref{sec:throughput_eval}). 
This sustained upward trend reflects {\system}’s ability to discover new cross-tensor delta compression opportunities even at scale, whereas ZipLLM plateaus at \rev{53.4\%} and standalone codecs---FM-Delta (37.1\%), ZipNN (33.3\%), and OpenZL (21.9\%)---remain flat without cross-model pairing.  
%\yuec{Fig10 is not described} 

Both {\system} variants remain robust because they do not rely on user-supplied model-card annotations.

\begin{revblock}
\phead{Attributing the Gains.}
The comparisons above separate deduplication, planning, and the codec.
\emph{Deduplication is never double-counted}: {\tensorid} first reduces 961,509 raw tensors (44.86\,TB) to 741,689 unique ones (40.11\,TB), and every ratio here---including the headline 70.5\%---is measured against that post-deduplication baseline.
\emph{Fixing the codec isolates the planner}: {\tensorx} is a byte-plane simplification of ZipLLM's BitX at a comparable ratio, so ZipLLM and TensorDex-TX differ only in how pairs are chosen, and the 53.4\%\,$\rightarrow$\,65.1\% gap is {\flexsplit}'s.
\emph{Fixing the plan isolates the codec}: swapping {\tensorx} for FM++ over the same pairs adds 5.4 points (65.1\%\,$\rightarrow$\,70.5\%) from its stronger entropy stage alone.
Planning is a one-time ingestion cost, absent from the retrieval path, and can be batched offline.
\end{revblock}

To better understand the source of these gains, we examine the per-tensor reduction distribution (\fref{fig:reduction_global_zipllm_trace}(b)). 
%As shown in the CDF, 
TensorDex-FM++ concentrates most tensors in the 60--90\% reduction range, while standalone codecs (FM-Delta, ZipNN) cluster at fixed reduction levels ($\sim$37\% and $\sim$33\%). 
%This contrast indicates that {\system}’s gains are not driven by a small number of highly compressible tensors, but instead arise consistently across the tensor population.
%
%Importantly, t
The strong performance of TensorDex-FM++ is not solely due to the underlying FM-Delta codec. Rather, it is {\system}’s tensor-centric planning---enabled by {\tensorsketch} and {\flexsplit}---that uncovers high-quality cross-tensor pairings, allowing even a strong standalone compressor to operate on substantially lower-entropy deltas. In contrast, when applied without {\system}’s planning, FM-Delta achieves only moderate, fixed compression gains. This demonstrates that the primary benefit comes from exposing previously hidden cross-tensor redundancy, with the compression codec further amplifying these gains.

% \begin{figure}[t]
%     \centering
%     \begin{subfigure}[t]{0.225\textwidth}
%         \centering
%         \includegraphics[width=\linewidth]{figures/evaluation/methods_bar.pdf}
%         \caption{Compressed size (\%) by methods.} 
%         %Average DRR by method combination.
%         %\tf{change to single column}}
%         \label{fig:bar_reduction_comparison}
%     \end{subfigure}
%     \hfill
%     \begin{subfigure}[t]{0.225\textwidth}
%         \centering
%         \includegraphics[width=\linewidth]{figures/evaluation/methods_cdf.pdf}
%         \caption{CDFs of tensor-level DRR.}   
%         \label{fig:cdf_methods_combination}
%     \end{subfigure}

%     \caption{Compression method comparison.
%     ({\bf a}) Normalized storage relative to uncompressed (Full = 1.00$\times$40.11TB).  
%     %: TensorDex-FM++ achieves 0.29$\times$ (71\% reduction), TensorDex-TX 0.35$\times$ (65\%), ZipLLM 0.47$\times$, FM-Delta 0.63$\times$, ZipNN 0.67$\times$, and OpenZL 0.78$\times$.
%     ({\bf b}) Per-tensor data reduction ratio (DRR) CDF.}
%     %: TensorDex-FM++ concentrates most tensors at 60--90\% reduction, while standalone codecs (FM-Delta, ZipNN) cluster at fixed rates ($\sim$37\% and $\sim$33\%), confirming that {\system}'s gains are consistent across the tensor population rather than driven by outliers.}
%     \label{fig:tensorx_ablation}
% \end{figure} 

\noindent\textbf{Per Family Compression Breakdown.}
To understand how compression effectiveness varies across different model lineages, we break down the results by ten representative base families.
\fref{fig:reduction_global_zipllm_trace}(c) reports the model-level reduction ratio for all fine-tuned derivatives of each family. 

Across families, TensorDex-FM++ consistently achieves the highest median reduction, followed by TensorDex-TX, with both substantially outperforming ZipLLM.
The largest gaps appear on Qwen2.5, Mistral-v0.1, and Gemma-3, where ZipLLM's reduction drops below 50\% while TensorDex-FM++ stays above 65\%---a 20-25 percentage-point advantage driven by {\system}'s tensor-level matching, which discovers better bases even when fine-tuned variants drift from the nominal base model.
For families like Qwen3 and Gemma-2, some individual models approach near-100\% reduction, reflecting highly homogeneous derivatives where all methods perform well; yet even here the TensorDex variants' wider violin ``bellies'' concentrate at higher ratios (75\%+), indicating more consistent compression across the population.

Overall, this per-family analysis demonstrates that {\system} is model-agnostic and robust regardless of the codec choice, providing large benefits when lineage structure is complex while still improving upon ZipLLM even for tightly clustered families.

\begin{table}[t]
\small
\centering
\caption{Data ingestion and retrieval throughput. \rev{Ingestion is end-to-end over the full pipeline (deduplication, fingerprinting, and codec encoding). All systems run with identical parallelism (192 threads), with all data in memory.}}
%; \tref{tab:storage_retrieval} measures retrieval on real storage backends.}}
\label{tab:compression_decompression_throughput}
% \scalebox{0.8}{
\begin{tabular}{lcc}
\toprule
\textbf{Method} & \textbf{Ingestion (MB/s)} & \textbf{Retrieval (MB/s)} \\
\midrule
ZipNN        & 1,473    & 9,575 \\
ZipLLM & 5,942     & 7,981 \\
%FM-Delta & 9,821     & 8,494  \\
OpenZL & 753     & 19,068  \\
\midrule
TensorDex-FM++ & 9,821 & 8,494 \\
TensorDex-TX & \textbf{22,916}  & \textbf{28,393} \\
\bottomrule
\end{tabular}
% }
\end{table}

%\begin{refine}
\subsubsection{Throughput Performance}
\label{sec:throughput_eval}

Table~\ref{tab:compression_decompression_throughput} reports end-to-end ingestion (deduplication + compression) and retrieval (decompression) throughput with 192 threads.
%TensorDex-TX achieves 22,916\,MB/s ingestion and 28,393\,MB/s retrieval, outperforming ZipLLM by $3.9\times$ and $3.6\times$, respectively.
%TensorDex-FM++ trades throughput for higher compression, 
%reaching XX,XXX\,MB/s ingestion and XX,XXX\,MB/s retrieval---still 
%while still remaining substantially faster than ZipLLM.

\noindent\textbf{Data Ingestion Throughput.}
{\system}’s ingestion pipeline consists of three stages---\rev{xxHash128} deduplication, {\tensorsketch} fingerprinting, and codec encoding---all implemented as SIMD-accelerated kernels over zero-copy \texttt{mmap} buffers (\sref{subsec:impl_operators}).
Because every stage is embarrassingly parallel across tensors and free of dynamic allocation, {\system} scales linearly with thread count.
TensorDex-TX sustains 22,916\,MB/s aggregate ingestion throughput thanks to {\tensorx}’s lightweight byte-plane pipeline.
TensorDex-FM++ trades throughput for higher compression, 
%reaching XX,XXX\,MB/s ingestion and XX,XXX\,MB/s retrieval---still 
while still remaining substantially faster than ZipLLM. 
%TensorDex-FM++ uses the enhanced FM-Delta++ codec, which we extended with \texttt{BF16} support and parallel execution to improve throughput from the original $\sim$100\,MB/s to 9,821\,MB/s.
%In contrast, ZipLLM’s BitX encoder incurs extra memory copies and unaligned accesses, limiting it to 5,942\,MB/s, while ZipNN reaches only 1,473\,MB/s.

\noindent\textbf{Data Retrieval Throughput.}
Retrieval cost is dominated by decompression, which is incurred on every model download.
TensorDex-TX’s {\tensorx} decompresses each chunk independently and recombines byte planes in parallel, sustaining 28,393\,MB/s---well above typical network and storage bandwidth, ensuring that decompression is never the retrieval bottleneck.
TensorDex-FM++ achieves 8,494\,MB/s, balancing decompression speed with its higher compression effectiveness. 
ZipLLM and ZipNN achieve 7,981\,MB/s and 9,575\,MB/s, respectively.

\begin{table}[t]
\small
\centering
\caption{\rev{End-to-end retrieval under real storage I/O, using the same model and TensorDex-TX decoder as \tref{tab:compression_decompression_throughput} and differing only in where the bytes come from. The \emph{uncompressed pull} rows are a no-delta bandwidth reference over the model's base blobs only.}}
\label{tab:storage_retrieval}
\begin{tabular}{llc}
\toprule
\textbf{Backend} & \textbf{Configuration} & \textbf{GB/s} \\
\midrule
\multicolumn{2}{l}{Codec kernel only (DRAM)} & 28.6 \\
\multicolumn{2}{l}{Full decode path (DRAM)}  & 7.9--8.8 \\
\midrule
\multirow{3}{*}{\makecell[l]{Object store\\(S3;\ 64\\in flight)}}
              & uncompressed pull       & 3.43 \\
              & cold, single-hop dedup  & 2.83 \\
              & warm bases              & \textbf{4.10} \\
\midrule
\multirow{4}{*}{\makecell[l]{Block device\\(EBS gp3,\\2000~MiB/s;\\16 in flight)}}
              & cold, uncompressed pull & 1.70 \\
              & cold, single-hop dedup  & 1.96 \\
              & cold, bases resident    & 2.92 \\
              & page-cache warm         & \textbf{4.26} \\
\bottomrule
\end{tabular}
\end{table}

\phead{Retrieval under Real Storage I/O.} 
\rev{We repeat the same retrieval with blobs fetched from real backends---Amazon S3 with parallel ranged GETs, and an NVMe-attached EBS gp3 volume read cold and warm---obtaining 1.96--4.26\,GB/s (\tref{tab:storage_retrieval}), 3--4.5$\times$ below the same decoder running on DRAM-resident blobs. Retrieval is therefore I/O-bound on both tiers, confirming that decompression is never the bottleneck. Reconstruction is one hop: each delta references exactly one base, so restoring a tensor reads only that base plus its delta, and read amplification stays at 1.2--1.4$\times$ (median 1.27$\times$ across the corpus). Caching popular bases can reduce it further.} 
%\end{refine}

% \begin{figure}[ht]
%     \centering 
%     \includegraphics[width=0.475\textwidth]{figures/background/source_diversity.pdf}
%     \caption{Optimal base tensors are scattered across many source models.
%     For each model (x-axis), we count the number of distinct source models from which its tensors' best delta base originates.
%     Most models draw their optimal bases from tens to over 30+ different source models.\yuec{not described}} 
%     \label{fig:source_diversity}
% \end{figure}

\begin{figure*}[t]
    \centering
    \includegraphics[width=0.95\textwidth]{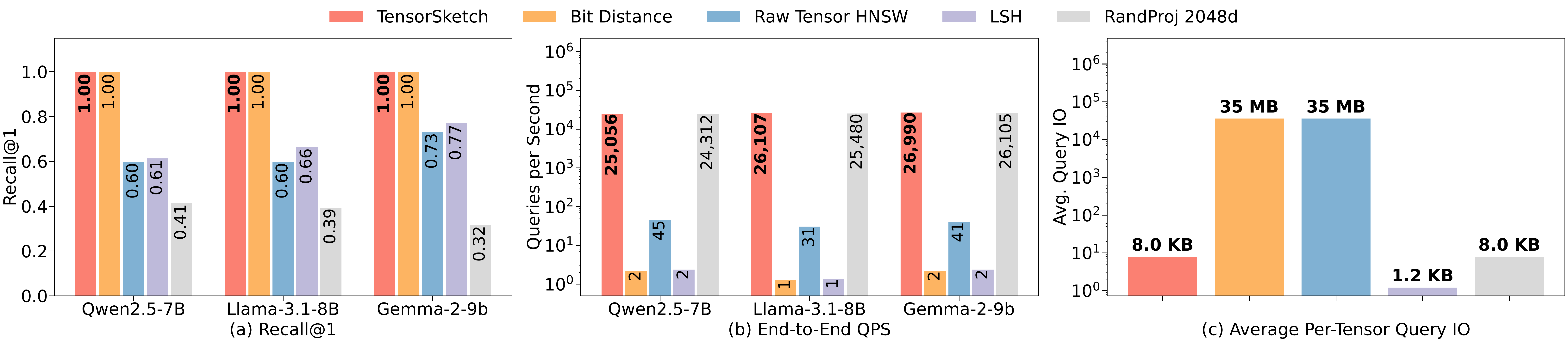}
    \caption{Performance Comparison of \tensorsketch. Comparison of \tensorsketch against baselines across Qwen, Llama, and Gemma families. ({\bf a}) \textbf{Recall@1} (top-1 match accuracy): \tensorsketch maintains a perfect 1.00 Recall@1, matching the exact Bit Distance baseline. ({\bf b}) \textbf{End-to-End QPS} (queries per second): \tensorsketch achieves over 25,000 QPS, representing a 4-order-of-magnitude speedup (up to 20,082$\times$) compared to Bit Distance. ({\bf c}) \textbf{Average Query IO}: \tensorsketch reduces the per-tensor query I/O from 35 MB (Bit Distance) to 8.0 KB, achieving a 4,517$\times$ reduction in IO overhead while maintaining significantly higher accuracy than other low-IO baselines like LSH.} 
    %(a)~Recall@1: BCS achieves $\sim$1.00 while baselines range 0.60--0.81.
    %(b)~Greedy reduction ratio: BCS maintains the highest reduction across all families.
    %(c)~End-to-end QPS (log scale): BCS sustains $>$25K~QPS, 500--800$\times$ faster than Raw Tensor HNSW.
    %(d)~Per-query data footprint: BCS reads $\sim$8\,KB vs.\ $\sim$35\,MB for Raw Tensor HNSW, a $>$4,000$\times$ reduction.}
    \label{fig:query}
\end{figure*}

\subsection{{\tensorsketch} and Prediction Model} 
\label{sec:sketch_pred_eval}

\phead{Similarity Search Baselines.}
To evaluate {\tensorsketch}'s effectiveness in identifying compressible tensor pairs, we compare against alternative similarity search methods paired with the same downstream planner:
\begin{itemize}[noitemsep,leftmargin=*]
    \item \textbf{Bit Distance~\cite{zipllm_nsdi26}:} computes the exact pairwise Hamming distance over raw tensor bytes via brute-force comparison.
    \item \textbf{Raw Tensor HNSW~\cite{HNSW2018}:} builds an HNSW index directly over raw tensor bytes, treating each full tensor as a high-dimensional vector for nearest-neighbor search.
    \item \textbf{LSH~\cite{LSH}:} applies locality-sensitive hashing to raw tensor representations, projecting tensors into hash buckets so that similar tensors are likely to collide.
    \item \textbf{RandProj 2048d~\cite{randproj,randproj_algo}:} applies a random linear projection (Johnson--Lindenstrauss) to reduce each tensor to a 2048-dimensional embedding, then builds an HNSW index over the projected vectors.
\end{itemize}

\phead{Sketch Method Comparison.}
We compare {\tensorsketch} (backed by HNSW, see \cref{subsec:impl_format}) against Raw Tensor HNSW, LSH, and RandProj 2048d across three major model families (Qwen, Llama, Gemma).
%We compare BCS HNSW ({\tensorsketch} backed by HNSW) against Raw Tensor HNSW, LSH, and RandProj 2048d across three major model families (Qwen, Llama, Gemma).
As shown in \fref{fig:query}, {\tensorsketch} dominates on all three metrics.

On \emph{Recall@1}, {\tensorsketch} achieves near-perfect recall ($\sim$1.00) across all families, while Raw Tensor HNSW and LSH range from 0.60 to 0.77 and RandProj 2048d from 0.32 to 0.41.
On \emph{greedy reduction ratio}, {\tensorsketch} consistently achieves the highest reduction, reaching approximately 60\% on the Gemma family, demonstrating that its high recall directly translates to better compression planning.
On \emph{end-to-end QPS}, {\tensorsketch} sustains over 25,000~QPS---500--800$\times$ faster than Raw Tensor HNSW ($\sim$31--45~QPS)---because its compact sketch reduces the per-query data footprint from $\sim$35\,MB (Raw Tensor HNSW) to only $\sim$8\,KB, a $>$4,000$\times$ reduction in I/O. 

\rev{The two approximations in this path---sketching, then HNSW---compound without cost, because all that matters for greedy attach is that the top-1 candidate stays stable. {\tensorsketch} matches exact bit distance at Recall@1 $=1.00$ (\fref{fig:query}(a)), and HNSW over those same sketches reaches Recall@1 $=0.9986$ against exact search, yielding an \emph{identical} {\flexsplit} outcome (65.1\% reduction, 6,482 bases) while clustering 3.8$\times$ faster. Sweeping \texttt{efSearch}
%$efSearch$ 
from 16 to 256 changes nothing.}

\begin{figure}[t]
    \centering
    \includegraphics[width=0.475\textwidth]{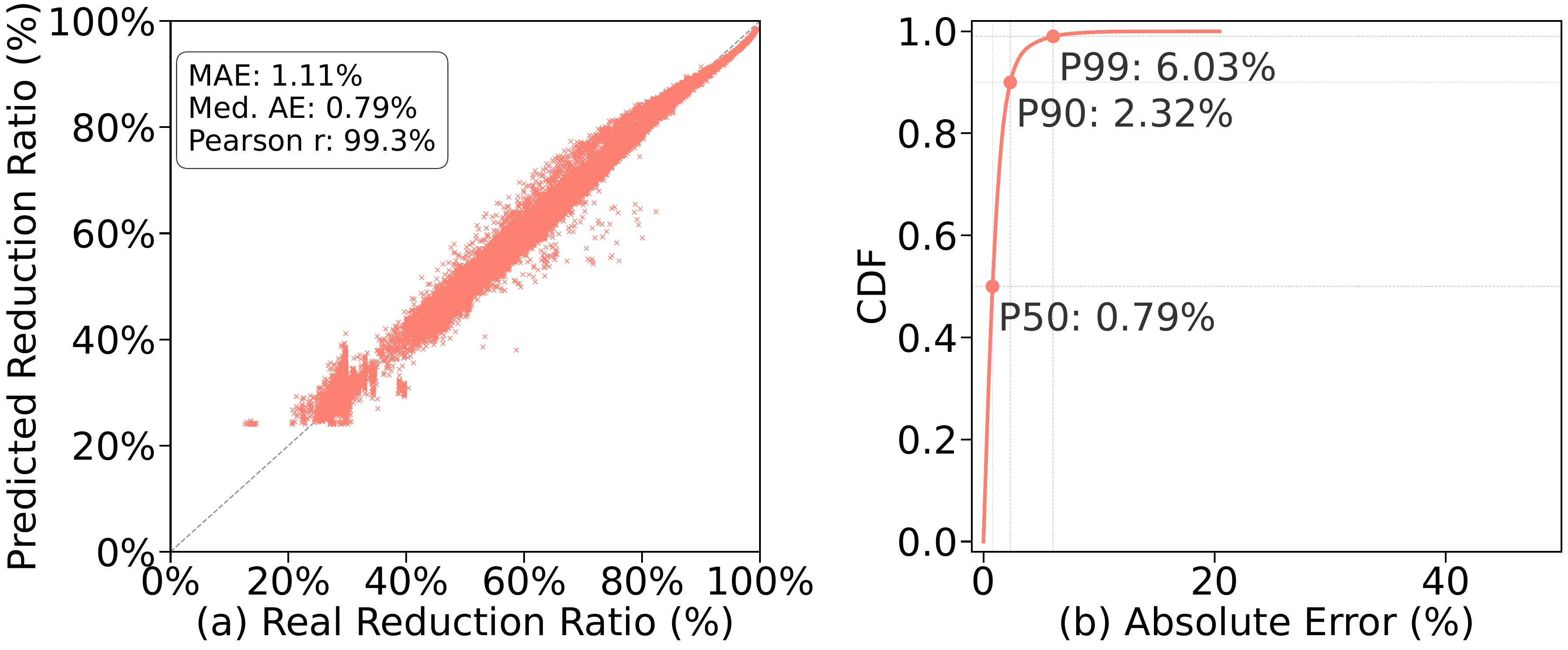}
    \caption{Validation of the reduction ratio prediction model.
    \rev{The coefficients $(\alpha,\beta,\gamma,\epsilon)$ come from a single global fit over 1M training pairs; the figure evaluates them on 5.7M \emph{held-out} validation pairs. \sref{sec:sketch_pred_eval} additionally reports a cross-family holdout.}
    ({\bf Left})~Predicted vs.\ real reduction ratio.  
    %(Pearson $r = 0.993$, MAE = 1.11\%, median error = 0.79\%).
    ({\bf Right})~CDF of absolute prediction error.} 
    %: P50 = 0.79\%, P90 = 2.32\%, P99 = 6.03\%, confirming that the lightweight predictor is sufficiently accurate for clustering and base-delta planning.}
    \vspace{1.6em}
    \label{fig:pred_model_val}
\end{figure}

\phead{Prediction Model Accuracy.}
Next, we evaluate the accuracy of the reduction prediction model. 
\fref{fig:pred_model_val}(a) plots predicted vs. real reduction ratio for 5.7M tensor pairs. The scatter tightly follows the diagonal (Pearson $r = 0.993$), with a mean absolute error of only 1.11\% and a median absolute error of 0.79\%. 
The close alignment across the full range---from 20\% to near-perfect reduction---shows that the lightweight sketch-based predictor faithfully captures delta compressibility without requiring full-tensor compression.

\fref{fig:pred_model_val}(b) presents the CDF of absolute prediction error. The model achieves a median error (P50) of 0.79\%, meaning that half of all predictions differ from ground truth by less than one percentage point. Even at the tail, errors remain small: P90 = 2.32\% and P99 = 6.03\%.
This level of accuracy is more than adequate for our use case: planning and clustering depend primarily on relative ordering among candidate bases rather than exact ratio magnitudes. The prediction model thus provides a lightweight yet effective mechanism for large-scale storage optimization.

\rev{To rule out family leakage, we refit on one set of families and test on disjoint ones: trained on Qwen$+$Llama (451K pairs) and evaluated on never-seen Gemma$+$Mistral (83K pairs), MAE rises only from 0.0230 to 0.0265 and Pearson $r$ is unchanged (0.978 vs.\ 0.977). The predictor's inputs are $\hat{p}$ and its binary entropy---numerical properties independent of model family---so one global regression fits all families.}

\subsection{{\salgo} Analysis}
\label{sec:flexsplit_eval}

%\phead{Optimal Solver.}
\phead{Evaluating Solvers.} 
To assess the scalability and quality of {\flexsplit}'s heuristic base selection, we compare against exact and classical approximation solvers:
\begin{itemize}[noitemsep,leftmargin=*]
    \item \textbf{ILP (Gurobi)~\cite{gurobi}:} an integer linear programming formulation solved by Gurobi, providing the global optimum but intractable beyond a few hundred tensors.
    \item \textbf{Primal-Dual~\cite{JainV01}:} the classical 3-approximation algorithm using the primal-dual schema. It offers provable guarantees but requires the full pairwise cost matrix and assumes static inputs.
\end{itemize}

\begin{figure}[t]
    \centering
    \begin{subfigure}[t]{0.225\textwidth}
        \centering
        \includegraphics[width=\linewidth]{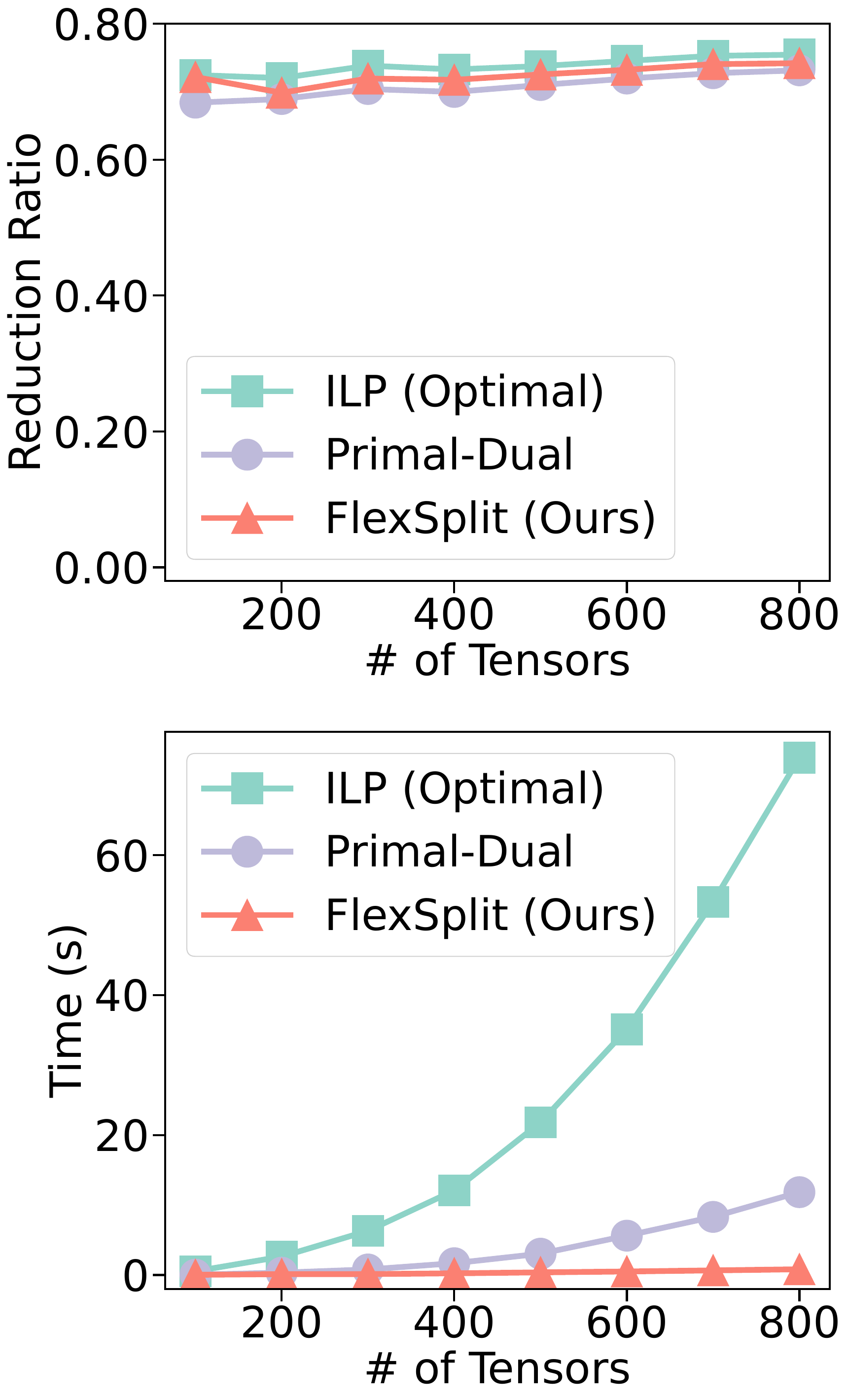}
        \caption{\texttt{q\_proj\_weight} tensors.}
        \label{fig:algo_benchmark_q_proj}
    \end{subfigure}
    \hfill
    \begin{subfigure}[t]{0.225\textwidth}
        \centering
        \includegraphics[width=\linewidth]{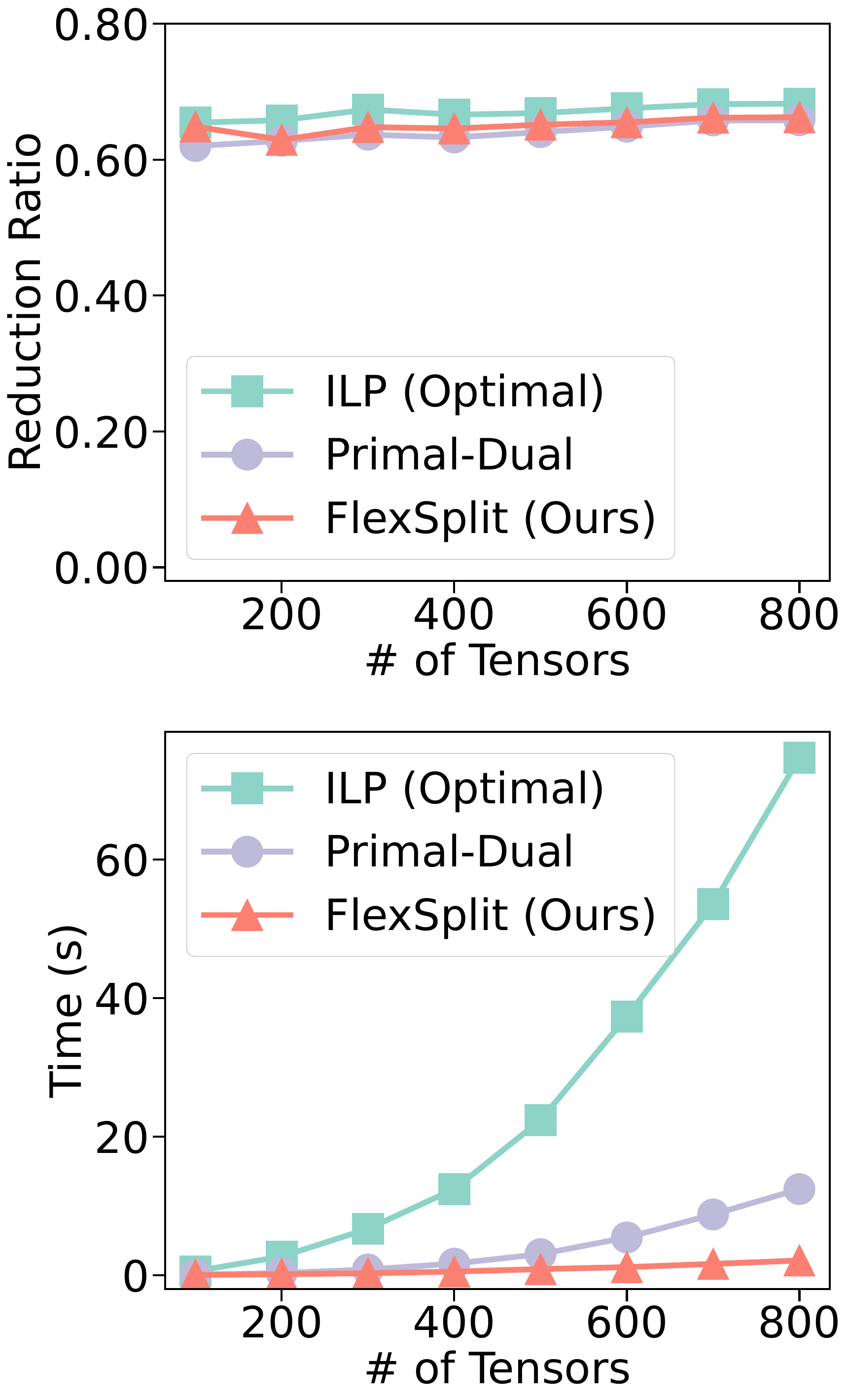}
        \caption{\texttt{v\_proj\_weight} tensors.}
        \label{fig:algo_benchmark_v_proj}
    \end{subfigure}

    \caption{Scalability of {\salgo} vs. ILP and Primal-Dual solvers on two representative tensor types.
    ({\bf Top})~Reduction ratio remains near-optimal for {\salgo} across all scales.
    ({\bf Bottom})~Solving time: ILP grows super-linearly and Primal-Dual grows linearly, while {\salgo} maintains near-constant time.}
    \label{fig:algo_scalability}
\end{figure}

\phead{Scalability.}
We benchmark the end-to-end solving time of {\salgo} against ILP and Primal-Dual over up to 800 \texttt{Qwen2.5-7B} tensors using two representative tensor types (\texttt{q\_proj\_weight} and \texttt{v\_proj\_weight}). 
The results highlight a sharp contrast in scalability (\fref{fig:algo_scalability}).
As the number of tensors grows from 100 to 800, ILP's solving time increases super-linearly (exceeding 70\,s at 800 tensors), making it impractical for production-scale model hubs.
Primal-Dual grows linearly but still increases 
%noticeably 
with scale. 
In contrast, {\salgo} maintains \emph{near-constant} solving time across all scales, approaching zero even at 800 tensors.
Its fingerprint-guided assignments and lightweight predictive splitting require only sketch-level distance evaluations, avoiding the expensive global cost computation that dominates the baselines.

Despite this vastly lower solve time, {\salgo} achieves a \emph{near-optimal} reduction ratio---stable at 0.60-0.75 and consistently within a few percentage points of the ILP optimum---demonstrating that the algorithm is both compute-efficient and high quality.
This scalability enables {\system} to continuously ingest large model families while retaining ILP-level compression effectiveness.

\begin{figure}[t]
    \centering
    \includegraphics[width=0.48\textwidth]{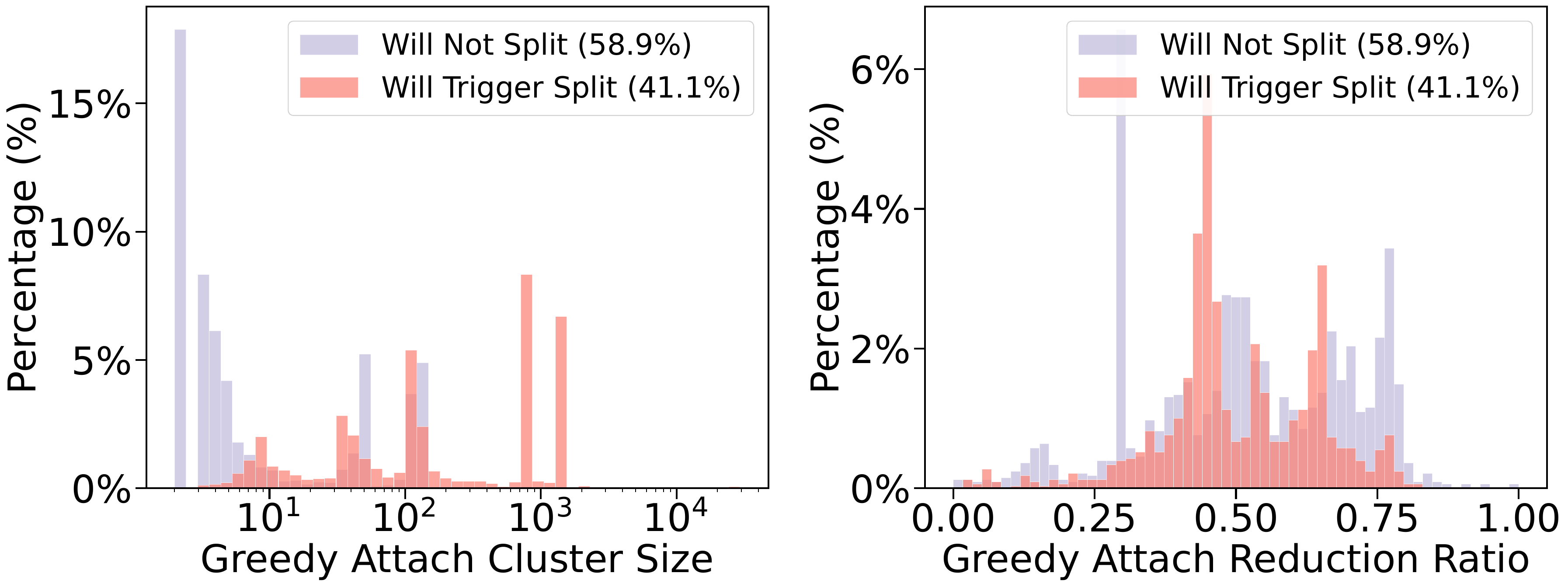}
    \caption{Cluster characteristics after Phase~I greedy assignment, categorized by whether Phase~II triggers a split.
    ({\bf Left})~Cluster size distribution (log-scale).
    %: non-split clusters concentrate at small sizes ($<$10), while split-targeted clusters dominate the $10^2$--$10^3$ range.
    ({\bf Right})~Reduction ratio distribution.}
    %: split-targeted clusters peak at 0.25--0.50, whereas non-split clusters peak near 0.75, confirming that Phase~II selectively refines the largest and worst-performing groupings.}
    \label{fig:greedy_attach}
\end{figure}

\phead{Splitting Behavior.}
We examine how {\salgo}'s two-phase structure behaves in practice over the full ZipLLM-Trace corpus.
\fref{fig:greedy_attach} characterizes the clusters produced by Phase~I greedy assignment.
The left panel shows cluster size distributions on a log scale: non-split clusters are concentrated at small sizes ($<$10 tensors), while split-targeted clusters (i.e., clusters that will trigger a split in Phase II) dominate the $10^2$--$10^3$ range, aggregating heterogeneous tensors from diverse fine-tuning lineages.
The right panel reveals a corresponding compressibility
%performance 
gap: split-targeted clusters peak at low reduction ratios (0.25--0.50), whereas non-split clusters peak near 0.75.
This confirms that Phase~II applies a dual criterion---it triggers a split only when a cluster is both sufficiently large and sufficiently inefficient to warrant refinement.
Overall, 58.9\% of clusters require no split at all, confirming that {\salgo}'s refinement is \emph{surgical} rather than pervasive.

\begin{figure}[t]
    \centering
    \includegraphics[width=0.48\textwidth]{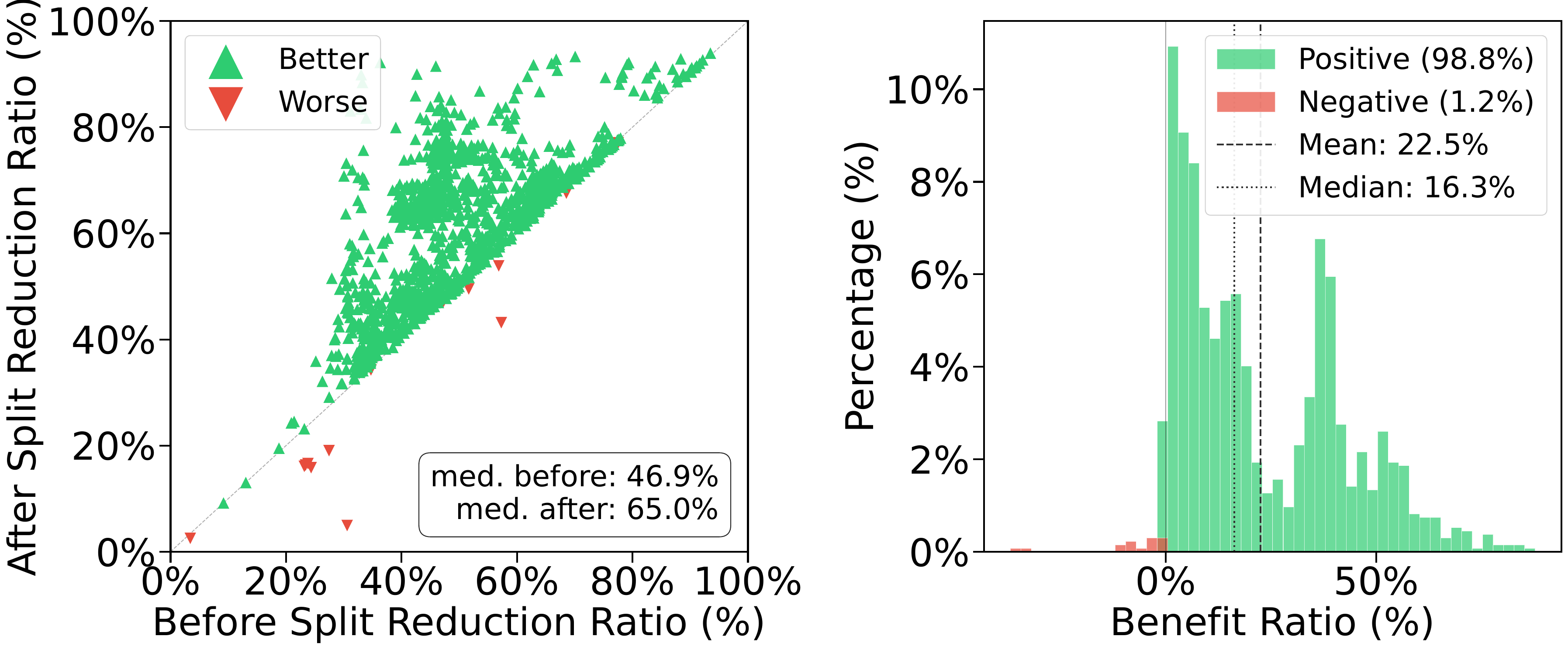}
    \caption{Effect of Phase~II {\salgo} splitting on per-cluster reduction ratio ($n = 1{,}352$ clusters).
    ({\bf Left})~Scatter of greedy assignment (Phase I) vs.\ {\salgo} (Phase II) reduction ratio.
    %: 98.8\% of split clusters improve (green), with median ratio increasing from 0.463 to 0.650; only 1.2\% show marginal regression (red, near zero).
    ({\bf Right})~Distribution of net gain ({\salgo} minus greedy).}   
    %: most gains fall between 0.1 and 0.3, confirming that Phase~II reliably adds 10--30\% additional reduction.}
    \vspace{1.6em}
    \label{fig:flexsplit_split_effect}
\end{figure}

\fref{fig:flexsplit_split_effect} quantifies the improvement from Phase~II splitting over 1,352 clusters.
The left panel plots per-cluster reduction ratio before (greedy assignment) vs. after ({\salgo}): 98.8\% of split clusters improve, with the median ratio increasing from 0.469 to 0.650. Only 1.2\% of cases show marginal regression, concentrated near zero.
The right panel shows the distribution of net gain: most improvements fall between 0.1 and 0.3, meaning Phase~II typically adds 10--30\% additional reduction per cluster.
This demonstrates that Phase~II reliably reassigns outlier tensors---those with poor fit to their initial base---to better-matched centers, directly translating into improved storage efficiency without accessing full tensor contents.

\section{Conclusion}
\label{sec:conclusion}

This paper presents {\system}, a compact, tensor-centric AI model storage system. 
%{\system} addresses the delta compression granularity mismatch of state-of-the-art model storage reduction method by exploiting data-driven cross-tensor relationships. 
{\system} addresses the granularity mismatch in existing delta compression approaches by exploiting data-driven cross-tensor relationships.
%Elevating tensors to the delta compression unit, {\system} 
{\system} synthesizes \rev{four} novel techniques:
%to minimize model storage redundancy. 
a compression-aware fingerprint that captures bit-level tensor similarity, a lightweight prediction model that estimates delta compressibility without accessing full data, \rev{an adaptive clustering planner that discovers high-quality base--delta tensor pairings at scale, and two pluggable high-performance delta codecs}. Together, these techniques enable an efficient, metadata-oblivious compression pipeline that substantially reduces storage while maintaining high performance in large-scale real-world model repositories.

\if 0
\draft{This paper presents {\system}, a novel tensor-centric model storage reduction system built on three key techniques: compression-aware fingerprinting and compressibility prediction (\tensorsketch and a regression-based prediction model), \salgo for scalable incremental clustering, and {\tensorx} for domain-specific delta compression of LLM weights.}
%unifies tensor-level deduplication and a new lossless delta compression called {\bitx} to address the growing scale of LLM storage. 
%Our large-scale study reveals key redundancies in LLM repositories and motivates design principles that synergize model storage deduplication with compression. 
\draft{{\system} achieves significantly higher storage savings compared to the state-of-the-art ZipLLM in real-world conditions, demonstrating that effective model storage reduction requires tensor-level pairing, metadata-free similarity discovery, and compression kernels tailored to the statistical structure of LLM weight deltas.}  
%, without sacrificing losslessness.  
%Looking ahead, our findings open promising opportunities for future research in LLM lineage tracking, file format optimization, and online quantization. 

%format optimization, online quantization, and structure-aware model storage.
\fi 
\section*{Acknowledgment}
\label{sec:acknowledgment} 

We thank our shepherd and the anonymous reviewers for their valuable feedback and comments, which improved the paper. This research was supported in part by NSF grants CNS-2322860, OAC-2411009, and OAC-2403313. 
We also acknowledge support from NSF CloudBank for providing AWS credits, thank Adobe for their generous research gift, and acknowledge support from  
the Commonwealth Cyber Initiative (CCI), an investment in the advancement of cyber research, innovation, and workforce development. For more information about CCI, visit \href{https://cyberinitiative.org}{cyberinitiative.org}. 

\bibliographystyle{plain}
\bibliography{reference}

% \newpage
% \input{sections/appendix}

\end{document}